\documentclass[aps,nofootinbib,showkeys,noshowpacs,preprintnumbers,amsmath,amssymb]{revtex4}
\pdfoutput=1

\def\be{\begin{equation}}
\def\ee{\end{equation}}
\def\ba{\begin{eqnarray}}
\def\ea{\end{eqnarray}}
\def\bea{\begin{eqnarray}}
\def\eea{\end{eqnarray}}
\def\bes{\begin{subequations}}
\def\ees{\end{subequations}}

\newcommand{\sm}{{\sigma_{\rm m}}}
\newcommand{\sg}{{\sigma}}
\newcommand{\tkap}{{\widetilde \kappa}}
\newcommand{\tk}{{\widetilde k}}
\newcommand{\tK}{{\widetilde K}}

\newcommand{\ta}{{\widetilde a}}
\newcommand{\tal}{{\widetilde \alpha}}

\newcommand{\td}{{\widetilde d}}

\newcommand{\tg}{{\widetilde {\gamma}}}
\newcommand{\bg}{{\overline {\gamma}}}
\newcommand{\MSbar}{\overline{\rm MS}}  

\usepackage{graphics}
\usepackage{graphicx}
\usepackage{dcolumn}
\usepackage{bm}
\usepackage{epsfig}
\usepackage{graphicx,color}
\usepackage{multirow}

\begin{document}


\title{Using improved Operator Product Expansion in Borel-Laplace Sum Rules with ALEPH $\tau$ decay data, and determination of pQCD coupling}

\author{C\'esar Ayala$^a$}
\email{c.ayala86@gmail.com}
\author{Gorazd Cveti\v{c}$^b$}
\email{gorazd.cvetic@gmail.com}
\author{Diego Teca$^b$}
\email{diegotecawellmann@gmail.com}
\affiliation{$^a$Instituto de Alta Investigaci{\'o}n, Sede Esmeralda, Universidad de Tarapac{\'a}, Av. Luis Emilio Recabarren 2477, Iquique, Chile}
\affiliation{$^b$Department of Physics, Universidad T{\'e}cnica Federico Santa Mar{\'\i}a (UTFSM), Casilla 110-V, Valpara{\'\i}so, Chile}

\date{\today}

\begin{abstract}
  We use improved truncated Operator Product Expansion (OPE) for the Adler function, involving two types of terms with dimension $D=6$, in the double-pinched Borel-Laplace Sum Rules and Finite Energy Sum Rules for the V+A channel strangeless semihadronic $\tau$ decays. The generation of the higher order perturbative QCD terms of the $D=0$ part of the Adler function is carried out using a renormalon-motivated ansatz incorporating the leading UV renormalon and the first two leading IR renormalons. The trunacted $D=0$ part of the Sum Rules is evaluated by two variants of the fixed-order perturbation theory (FO), by Principal Value of the Borel resummation (PV), and by  contour-improved perturbation theory (CI). For the experimental  V+A channel spectral function we use the ALEPH $\tau$-decay data. We point out that the truncated FO and PV evaluation methods account correctly for the renormalon structure of the Sum Rules, while this is not the case for the truncated CI evaluation. We extract the value of the ${\overline {\rm MS}}$ coupling $\alpha_s(m_{\tau}^2) = 0.3235^{+0.0138}_{-0.0126}$ [$\alpha_s(M_Z^2)=0.1191 \pm 0.0016$] for the average of the two FO methods and the PV method, which we consider as our main result. If we included in the average also CI extraction, the value would be  $\alpha_s(m_{\tau}^2) = 0.3299^{+0.0232}_{-0.0225}$ [$\alpha_s(M_Z^2)=0.1199^{+0.0026}_{-0.0028}$]. This work is an extension and improvement of our previous work \cite{EPJ21} where we used for the truncated OPE a more naive (and widely used) form and where the extracted values for $\alpha_s(M_Z^2)$ were somewhat lower.
 \end{abstract}
\keywords{perturbative QCD; QCD phenomenology; semihadronic $\tau$ decays; renormalons}

\maketitle


\section{Introduction}
\label{sec:Int}

One of the most important challenges of physics is the determination of fundamental parameters. In the case of strong interactions, the main parameter is the strong running coupling $a(Q^2) \equiv \alpha_s(Q^2)/\pi$ which depends on the squared (spacelike) momentum scale $Q^2 (\equiv - q^2)$ characteristic of the process. This coupling is well determined at high energies $1 \ {\rm GeV}^2 < Q^2$ ($ \lesssim M_{Z}^2$) because of the high precision of the experiments and the theory. On the other hand, determinations at lower energies are a good test of the consistency of the theory. We study here the quantities related with the semihadronic decay width of $\tau$ lepton, which is at low scales $Q \lesssim m_{\tau}$ ($\sim 1$ GeV) and for which high precision experimental results are available from the ALPEH collaboration \cite{ALEPH2,DDHMZ,ALEPHfin,ALEPHwww}. Both the experimental and the theoretical results are related with the two-point correlation function of ($u d$) quark currents $\Pi(Q^2)$; the experimental results are given in terms of the spectral function $\omega(\sigma) \propto {\rm Im} \Pi(-\sigma - i \epsilon)$, while the theoretical results are usually given in terms of contour integrals of the derivative of the correlation function, ${\cal D}(Q^2)\propto d \Pi(Q^2)/d \ln Q^2$, known as the Adler function.

Specific contour integrals involving the Adler function give us various $\tau$-decay sum rules, including the inclusive strangeless decay ratio $R_{\tau} = \Gamma(\tau \to \nu_{\tau} {\rm hadrons})/\Gamma(\tau \to \nu_{\tau} e^- \bar{\nu}_{e})$ \footnote{\label{rtau} The QCD part (canonical part) of $R_{\tau}$ is the quantity $r_\tau=r_\tau^{(D = 0)} + \delta r_\tau(m_{u, d} \not= 0) + \sum_{D \geq 4} r_{\tau}^{(D)}$ and it appears in the semihadronic strangeless V+A $\tau$-decay ratio $R_{\tau}$ via the relation $R_{\tau}= 3 |V_{ud}|^2 S_{\rm EW} ( 1 + \delta'_{\rm EW} + r_{\tau})$, where $S_{\rm EW}=1.0198 \pm 0.0006$ \cite{SEW} and $\delta r_\tau(m_{u,d} \not= 0) \approx - 8 \pi^2 f_\pi^2 m_\pi^2/m_\tau^4 \approx -0.0026$ (with $f_{\pi}=0.1305$ GeV), cf.~\cite{NP88,B88B89,BNP92,GCTK01}, $\delta'_{\rm EW}=0.0010 \pm 0.0010$ \cite{dpEW} are electroweak corrections, $V_{ud}$ is the CKM matrix element.}. The Adler function was calculated by perturbation techniques of QCD up to ${\cal O}(\alpha_s^4)$, cf.~\cite{d1,d2,BCK}. Determinations of the coupling from such sum rules, i.e., at low momenta $\sim m_{\tau}$, allow us to see the reliability of the theory because we can compare the extracted value of $\alpha_s$  with the known determinations at high energies through renormalisation group equation (RGE) evolution (see for instance \cite{DBTrev,alpha2019,PDG2020}).

For the theoretical expressions of the sum rules, the Operator Product Expansion (OPE) is used for the quark current correlator (and the related Adler function). This implies that the sum rules have the perturbative part (dimension $D=0$), and the nonperturbative corrections ($D >0$), and the latter are often small such as in the case of $R_{\tau}$ ($r_{\tau}$) \cite{BNP92,DP92}. One of the novel aspects covered in the present work, in comparison with the previous one \cite{EPJ21}, is that we will deal with the nonperturbative contributions more carefully. In particular here the $D$ ($\equiv 2 k$) $ \geq 6$ condensate contributions to the Adler function $d(Q^2)$ have two parts, which eliminate the renormalon ambiguity originating from the two infrared (IR) renormalon contributions of the Adler function at $u = k$ ($k=3,4,\ldots$), one of the type $\sim 1/(k - u)^{\tg_k+1}$ and the other $\sim 1/(k-u)^{\tg_k}$ (where $\tg_k= 1 + k \beta_1/\beta_0^2$ is unity in the large-$\beta_0$ limit).
\be
d(Q^2)_{D=2 k} = 2 \pi^2 k \left[ \frac{\langle {\bar O}^{(2)}_{2 k} \rangle}{a(Q^2) (Q^2)^k} + \frac{\langle {\bar O}^{(1)}_{2 k} \rangle}{(Q^2)^{k}} \right].
\label{dD2k} \qquad (k=3,4,\ldots). \ee
The $D>6$ condensates will not be included in our analysis, as we use an extended Adler function that is based on a renormalon-motivated ansatz that includes only the first two IR renormalons $u=2,3$ (and the first UV renormalon $u=-1$) \cite{renmod}.

Evaluation of the perturbative part ($D=0$) of sum rules is based on (re)summations and truncations. These (re)summations require integration of the perturbation series on a circular contour in the complex $Q^2$-plane with radius $|Q^2|=\sigma_{\rm max}$ ($\sim m^2_{\tau}$), which involve the QCD running coupling parameter $a(Q^2)$ along this circle. Due to truncations, different ways of evaluation give different results. We will apply four of them. The first two are the fixed-order (FO) approach and its variant ${\widetilde {\rm FO}}$, where the coupling in the Adler function is Taylor-expanded around the spacelike point $Q^2=\sigma_{\rm max} >0$, and in the FO the series in powers of $a(\sigma_{\rm max})$ is performed and truncated, while in the ${\widetilde {\rm FO}}$ the series in logarithmic derivatives of $a(\sigma_{\rm max})$ is performed and truncated. The other is the contour-improved (CI) evaluation, which evolves $a(Q^2)$ through RGE along the contour $Q^2=\sigma_{\rm max} e^{i \phi}$. The third is the so called principal-value (PV) evaluation, where the inverse Borel transformation is appplied to the singular part of the Borel transform of the Adler function and the principal value prescription is applied for the integration across the IR renormalon singularities \cite{CI1,CI2,CIAPT}.

The extraction of $\alpha_s(m_\tau^2)$ from the $\tau$-decay data was performed in the past in the literature with the FO and CI approaches, giving different results. The tendency shows that the CI gives higher values of $\alpha_s(m_\tau^2)$ than the FO, even if the duality violation effects are explicitly taken into account  with a model \cite{Cata,BCK,ALEPHfin,Pich}. The discrepancy between the CI and FO was discussed in \cite{BJ,BJ2} in the large-$\beta_0$ (LO) approximation and then beyond-LO (bLO) in a renormalon-motivated model. The authors of \cite{BJ,BJ2} argued that the truncated FO approach in the $r_{\tau}$ sum rule takes correctly into account some cancellations of the leading renormalon contributions of the Adler function. They presented theoretical arguments for this cancellations in the LO approximation. Further, they argued that such cancellation do not take place in the truncated CI approach. Beyond LO (bLO), such cancellations were demonstrated in an approach with a modified Borel transform in a specific renormalisation scheme \cite{BoiOl}. In our previous work \cite{EPJ21}, we presented arguments that such cancellations take place in sum rules at the bLO level in any renormalisation scheme when the $D=0$ contribution to the sum rule is written in terms of the series in logarithmic derivatives of $a(\sigma_{\rm max})$ (cf.~Appendix  of \cite{EPJ21}).

The main goal of this work is to determine the value of the running coupling $\alpha_s(m_\tau^2)$, by using a renormalon-motivated extension \cite{renmod} of the Adler function for the $D=0$ contribution and the OPE with the terms $D=4$ and $D=6$ Eq.~(\ref{dD2k}). The sum rules we use for this determination are the Borel-Laplace sum rules, which are in the double-pinched form in order to suppress the duality-violating effects.\footnote{Double-pinched forms of sum rules were shown to suppress sufficienty well the duality-violating effects in the work of \cite{Pich} (cf.~also \cite{BNP92,DP92,Chib,Malt,DomSch,Cir,GonzAl,Dom,RSan}).} The $D=0$ contribution is evaluated by the (truncated) methods FO, ${\widetilde {\rm FO}}$, PV and CI. The truncation index of these contributions is then determined by considering the double-pinched finite energy sum rules (FESRs) $a^{(2,0)}$ and $a^{(2,1)}$ and finding the value of the truncation index for which these quantities become locally stable under the variation of the index. This work can be regarded as a continuation and improvement of our previous analysis \cite{EPJ21}, where now the improved form (\ref{dD2k}) of the OPE terms is used; further, the variation of the higher order perturbation coefficients of the renormalon-motivated extension of the Adler function is performed now in a more realistic way.

This paper is organized as follows. In Sec.~\ref{sec:SRgen} we present the improved $D \geq 6$ terms of the OPE for the Adler function, and summarise the general form of the sum rules for the (strangeless) semihadronic $\tau$-decay data. In Sec.~\ref{sec:renmod} we present briefly the renormalon-motivated extension of the $D=0$ Adler function, including the variation of the parameters of the model which reflect the variation of the higher order perturbation coefficients. In Sec.~\ref{sec:SR} the specific sum rules used later in the analysis are presented. In Sec.~\ref{sec:methD0} we summarise the various (truncated) methods used for the evaluation of the $D=0$ contribution to the sum rules. Section \ref{sec:res} is the main part of this work, it contains the numerical analysis of the extraction of the value of $\alpha_s(m^2_{\tau})$ and of the $D=4, 6$ condensates, for the full V+A channel. In Sec.~\ref{sec:concl} we summarise the results and present our conclusions. \textcolor{black}{A summary of our results is given in \cite{Trento}.}

\section{Sum rules and Adler function}
\label{sec:SRgen}

The Adler function ${\cal D}(Q^2)$ is defined as a logarithmic derivative of the quark current polarisation function $\Pi(Q^2)$
\be
{\cal D}(Q^2) \equiv - 2 \pi^2 \frac{d \Pi(Q^2)}{d \ln Q^2} .
\label{Ddef}
\ee   
We will consider the total (V+A)-channel, i.e., $\Pi(Q^2)$ will be the total (V+A)-channel polarisation function
\be
\Pi(Q^2) = \Pi_{\rm V}^{(1)}(Q^2) + \Pi_{\rm A}^{(1)}(Q^2) + \Pi_{\rm A}^{(0)}(Q^2).
\label{Pidef}
\ee
In this V+A sum (\ref{Pidef}), the term $\Pi_{\rm V}^{(0)}(Q^2)$ gives negligible contribution (to the sum rules) because ${\rm Im} \Pi_{\rm V}^{(0)}(-\sigma + i \epsilon) \propto (m_d-m_u)^2$. Further, we will neither include corrections ${\cal O}(m^2_{u,d})$ and  ${\cal O}(m^4_{u,d})$ for being numerically negligible.
The functions $\Pi_{J}^{(i)}$ ($J=V, A$ and $i=0,1$) characterise the quark current correlator
\be
\Pi_{\rm{J}, \mu\nu}(q) =  i \int  d^4 x \; e^{i q \cdot x} 
\langle T J_{\mu}(x) J_{\nu}(0)^{\dagger} \rangle
=  (q_{\mu} q_{\nu} - g_{\mu \nu} q^2) \Pi_{\rm J}^{(1)}(Q^2)
+ q_{\mu} q_{\nu} \Pi_{\rm J}^{(0)}(Q^2),
\label{PiJ}
\ee
where $J_{\mu}$ are the up-down quark vector and axial vector currents, $J_{\mu} = {\bar u} \gamma_{\mu} d$ and ${\bar u} \gamma_{\mu} \gamma_5 d$ for $J=V, A$, respectively. We recall that $q^2 \equiv -Q^2$ is the square of the momentum transfer, $q^2=(q^0)^2 - {\vec q}^2$.  

\begin{figure}[htb] 
\centering\includegraphics[width=70mm]{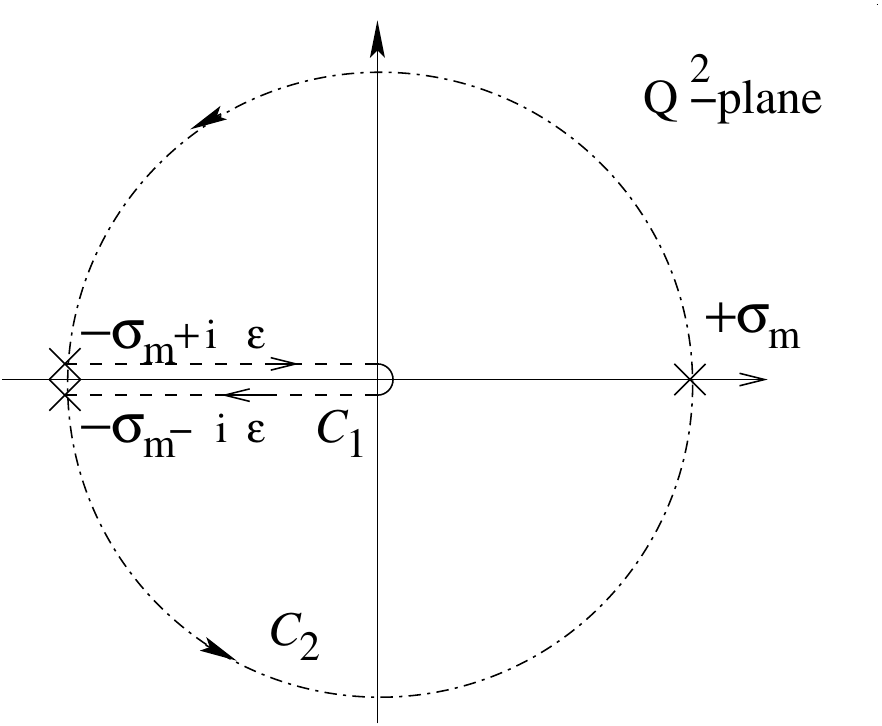}
\caption{\footnotesize The closed integration path $C_1+C_2$ for $\oint d Q^2 g(Q^2) \Pi(Q^2)$. The radius of the circle $C_2$ is $|Q^2|=\sigma_{\rm max}$ ($\equiv \sm$) ($\leq m_{\tau}^2$). In the sectors of the path $C_1$ we have $\varepsilon \to +0$.}
\label{Figcont2}
 \end{figure}

\begin{figure}[htb] 
  \centering\includegraphics[width=110mm]{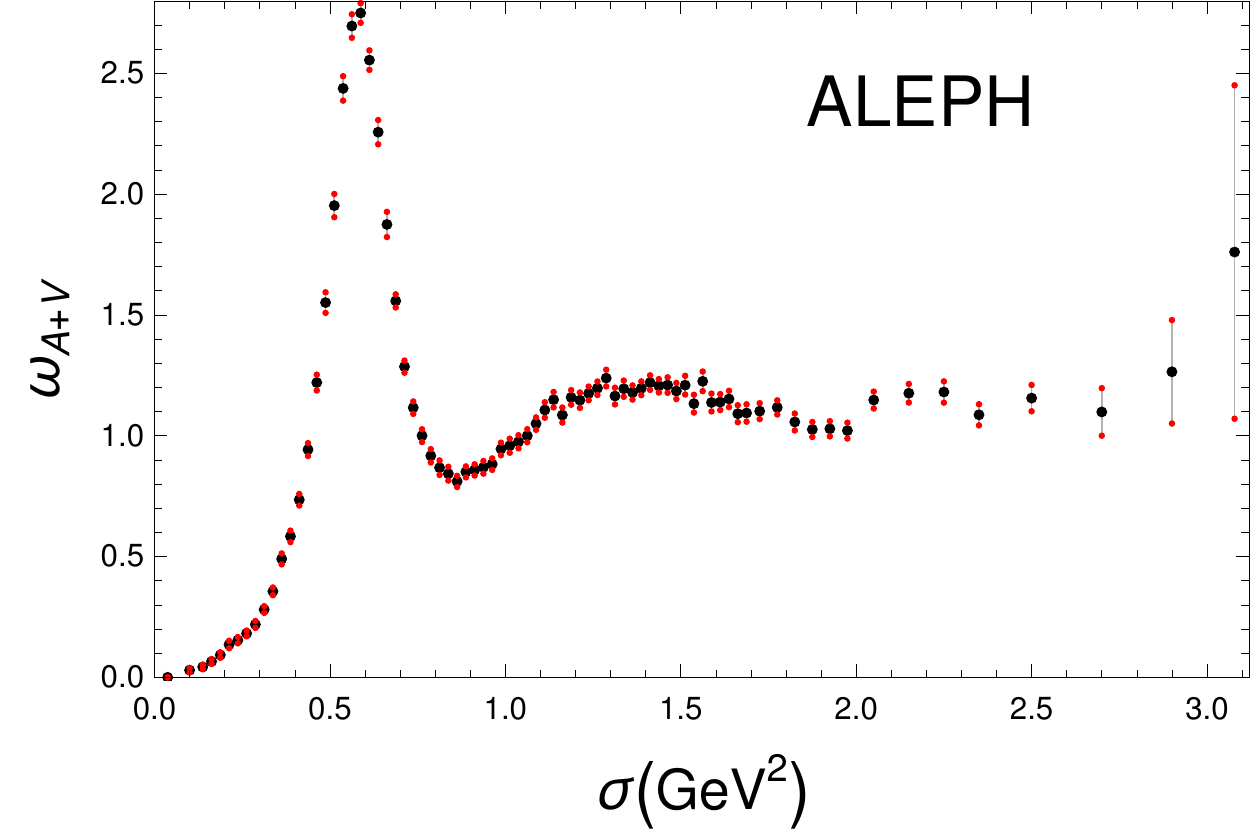}
\vspace{-0.2cm}
\caption{\footnotesize  (coloured online) The spectral function $\omega(\sigma)$ for the (V+A)-channel, as measured by ALEPH Collaboration \cite{ALEPH2,DDHMZ,ALEPHfin,ALEPHwww}. The extremely narrow pion peak contribution $2 \pi^2 f^2_{\pi} \delta(\sigma - m^2_{\pi})$ ($f_{\pi}=0.1305$ GeV) has to be added to this. The last two bins have large uncertainties, so we exclude them, and this means that $\sm =2.80 \ {\rm GeV}^2$ in the sum rules.}
\label{FigOmega}
\end{figure}

The usually used theoretical expression of the polarisation function has the following OPE \cite{SVZ} form:
\be
\Pi_{\rm th}(Q^2; \mu^2) = - \frac{1}{2 \pi^2} \ln \left( \frac{Q^2}{\mu^2} \right) + \Pi(Q^2)_{D=0} + \sum_{k \geq 2} \frac{ \langle O_{2 k} \rangle}{(Q^2)^k} \left( 1 + {\cal O}(a) \right).
\label{PiOPE1}
\ee
Here, $\mu^2$ is the squared renormalisation scale, and $\langle O_{2 k} \rangle \equiv \langle O_{2 k} \rangle_{\rm V+A}$ are condensates (vacuum expectation values) of dimension $D=2 k$ ($\geq 4$), for the full channel V+A. Such OPE form is usually used in numerical analyses, cf.~\cite{Pich,Pich3,RodS,Bo2011,Bo2012,Bo2015,Bo2017,EPJ21}. The corresponding Adler function (\ref{Ddef}) is then
\be
{\cal D}_{\rm th}(Q^2) \equiv  - 2 \pi^2 \frac{d \Pi_{\rm th}(Q^2)}{d \ln Q^2} = 1 + d(Q^2)_{D=0} + 2 \pi^2 \sum_{k \geq 2} \frac{k \langle O_{2 k} \rangle}{ (Q^2)^k}.
\label{DOPE1}
\ee
The $D=2 k$ term in this expansion is related to a renormalon singularity of the Borel transform ${\cal B}[d](u)$ of the Adler function $d(Q^2)_{D=0}$. Namely, if the perturbation expansion of $d(Q^2)_{D=0}$ in powers of $a(\mu^2) \equiv \alpha_s(\mu^2)/\pi$ is\footnote{Here, $\kappa \equiv \mu^2/Q^2$ ($0 <\kappa \sim 1$) is the dimensionless parameter for the renormalisation scale $\mu^2$.} 
\be
d(Q^2)_{D=0, {\rm pt}}= d_0 a(\kappa Q^2) + d_1(\kappa) \; a(\kappa Q^2)^2 + \ldots + d_n(\kappa) \; a(\kappa Q^2)^{n+1} + \ldots, \qquad (d_0=1),
\label{dpt}
\ee
the expansion of the Borel transform ${\cal B}[d](u; \kappa)$ is
\be
{\cal B}[d](u; \kappa) \equiv d_0 + \frac{d_1(\kappa)}{1! \beta_0} u + \ldots + \frac{d_n(\kappa)}{n! \beta_0^n} u^n + \ldots \ .
\label{Bdexp} \ee
This Borel transform has singularities at positive $u=2,3, \ldots$ [infrared (IR) renormalons] and at negative $u=-1,-2,\ldots$ [ultraviolet (UV) renormalons]. It turns out that the $D= 2 k$ term in the OPE expansion (\ref{DOPE1}) has the form which cancels the ambiguity coming from the infrared renormalon singularity at the value $u=k$ of the Borel transform ${\cal B}[d](u)$ of the Adler function $d(Q^2)_{D=0}$. However, this cancellation occurs only if the renormalon singularity has the form $\sim 1/(k - u)^{\tg_k}$, where\footnote{The $\beta_j$ coefficients in our convention appear in the following form of the RGE for $a(Q^2)$: $d a(Q^2)/d \ln Q^2 = - \beta_0 a(Q^2)^2 - \beta_1 a(Q^2)^3 - \beta_2 a(Q^2)^4 - \ldots$. The one-loop and two-loop coefficients $\beta_0$ and $\beta_1$ are universal (i.e., scheme independent) in mass independent schemes, $\beta_0 = (11 - 2 N_f/3)/4$ ($=9/4$ for $N_f=3$) and $\beta_1=(1/16)(102 - 38 N_f/3)$. The other coefficients ${\beta}_j$ ($j \geq 2$) depend on (and characterise) the renormalisation scheme.}$^{,}$\footnote{In the general case, the power in $\sim 1/(k -u)^{\tg_k}$ is $\tg_k =1 + k \beta_1/\beta_0^2 - \gamma_{O_{D}}^{(1)}/\beta_0$, where $\gamma_{O_{D}}^{(1)}$ is the effective leading-order anomalous dimension of the $D$-dimensional OPE operator $O_D$ ($D=2 k$), and the corresponding term in the OPE (\ref{DOPE1}) would have in such a case, instead of $\langle O_D \rangle$, the expression $a(Q^2)^{\gamma_{O_{D}}^{(1)}/\beta_0} \langle O_D \rangle$ \cite{renmod} [cf.~also the later discussion in the second paragraph after Eq.~(\ref{tdnkap})].} $\tg_k =1 + k \beta_1/\beta_0^2$. In the leading-$\beta_0$ (LB) approximation (i.e., where $\beta_1=0$), we have $\tg_k=1$, i.e., the singularity $\sim 1/(k - u)^{\tg_k}$ corresponds in the LB approximation to a simple single pole. However, the Borel transform of the $D=0$ Adler function is known to all orders in the LB approximation \cite{LB1,LB2,ren}, and it turns out that only the $u=2$ singularity is a single pole, but all other IR singularities (at $u=3, 4, \ldots$) are combinations of double and single poles. Consequently, when going beyond the LB approximation, the renormalon poles of the Adler function at $u=k \geq 3$ become $\sim 1/(k - u)^{\tg_k+1}$ and $\sim 1/(k - u)^{\tg_k}$ (and lower singularities), \textcolor{black}{if we assume that the beyond-LB effects come from the RGE-evolution (beyond one-loop) of the QCD coupling and that the leading-order anomalous dimensions of the corresponding operators remain of the LB-type.}\footnote{\textcolor{black}{Cf.~discussion in the next Section on this point.}} The corresponding OPE terms which cancel the ambiguities from these singularities at $u=k$ ($k \geq 3$) are then $\sim 1/(Q^2)^k/a(Q^2)$ and $1/(Q^2)^k$ \cite{renmod}.\footnote{This can also be formulated in the following way \cite{renmod}: the corresponding dimension $D \equiv 2 k$ operators have the leading-order anomalous dimension coefficient $-\gamma^{(1)}_{O_D}/\beta_0=+1, 0$, respectively.} This implies that the OPE expansion of the Adler function, \textcolor{black}{improved with respect to the expansion (\ref{DOPE1}),} has the following form: 
\be
   {\cal D}_{\rm th}(Q^2) \equiv  - 2 \pi^2 \frac{d \Pi_{\rm th}(Q^2)}{d \ln Q^2} = d(Q^2)_{D=0} + 1 + 4 \pi^2 \frac{\langle O_{4} \rangle}{ (Q^2)^2} + 2 \pi^2 \sum_{k \geq 3} \frac{k}{ (Q^2)^k} \left[ \frac{\langle O_{2 k}^{(2)} \rangle}{a(Q^2)} + \langle O_{2 k}^{(1)} \rangle \right],
\label{DOPE}
\ee
which has two different condensates for each $D = 2 k \geq 6$. This form of the OPE terms was noted already in our previous work \cite{EPJ21} [Eq.~(59) there], but the implementation was left pending. In the expansion (\ref{DOPE}), we neglected terms $O(a)$, i.e., terms $\sim \langle O_{2 k} \rangle a(Q^2)$, because they are relatively small for the considered momenta ($|Q^2| \approx 3 \ {\rm GeV}^2$). 
It can be checked that this expansion then corresponds to the following expansion of the polarisation function:
\bea
\Pi_{\rm th}(Q^2; \mu^2) &=& - \frac{1}{2 \pi^2} \ln \left( \frac{Q^2}{\mu^2} \right) + \Pi(Q^2)_{D=0} + \frac{ \langle O_4 \rangle }{(Q^2)^2} (\left( 1 + {\cal O}(a) \right)
\nonumber\\ &&
+ \sum_{k \geq 3} \frac{1}{(Q^2)^k} \left[ (1 - \beta_0) \frac{\langle O_{2 k}^{(2)}\rangle}{a(Q^2)} + \langle O_{2 k}^{(1)} \rangle \left( 1 + {\cal O}(a) \right) \right].
\label{PiOPE}
\eea
We will use this OPE expansion (\ref{DOPE})-(\ref{PiOPE}) in the sum rules. According to the general principles of Quantum Field Theory, the considered polarisation function $\Pi(Q^2;\mu^2)$ and its logarithmic derivative ${\cal D}(Q^2)$ are holomorphic (i.e., analytic) functions of $Q^2$ in the complex $Q^2$-plane with the exception of the real negative axis $(-\infty, - m^2_{\pi})$. Then if $g(Q^2)$ is a (arbitrary) holomorphic function of $Q^2$, then the Cauchy theorem can be applied to the integral $\oint dQ^2 g(Q^2) \Pi(Q^2;\mu^2)$ along a closed path in the complex $Q^2$-plane that consists, for example, of the circle of finite radius $|Q^2| = \sigma_{\rm max}$ ($\equiv \sm$) and lines above and below the negative axis avoiding thus the enclosure of the values $Q^2<0$ where the integrand is not holomorphic (cf.~Fig.~\ref{Figcont2}).
This then implies
\bes
\label{Cauchy}
\bea
\oint_{C_1+C_2} d Q^2 g(Q^2) \Pi(Q^2) & = & 0  
\label{Cauchya} \\
\Rightarrow \;\; \int_0^{\sm} d \sigma g(-\sigma) \omega_{\rm exp}(\sigma)  &=&
-i \pi  \oint_{|Q^2|=\sm}
d Q^2 g(Q^2) \Pi_{\rm th}(Q^2) ,
\label{Cauchyb} \eea \ees
where the integration on the right-hand side of Eq.~(\ref{Cauchyb}) is counterclockwise ($Q^2 = \sm e^{i \phi}$, $-\pi < \phi < \pi$), and $\omega(\sigma)$ is proportional to the discontinuity (spectral) function of the (V+A)-channel polarisation function
\be
\omega(\sigma) \equiv 2 \pi \; {\rm Im} \ \Pi(Q^2=-\sigma - i \epsilon) \ .
\label{om1}
\ee
The quantity $\omega(\sigma)$ was measured by OPAL Collaboration \cite{OPAL,Bo2012}, and to an even higher precision by ALEPH Collaboration \cite{ALEPH2,DDHMZ,
ALEPHfin,ALEPHwww}, in semihadronic strangeless decays of the $\tau$ lepton. In our analysis we will use the ALEPH data; they are presented in Fig.~\ref{FigOmega}.
Integration by parts allows us to replace the theoretical polarisation function in the sum rule (\ref{Cauchyb}) by the Adler function (\ref{Ddef})
\be
\int_0^{\sm} d \sigma g(-\sigma) \omega_{\rm exp}(\sigma)  =
\frac{1}{2 \pi}   \int_{-\pi}^{\pi}
d \phi \; {\cal D}_{\rm th}(\sm e^{i \phi}) G(\sm e^{i \phi}) ,
\label{sr}
\ee 
where  ${\cal D}_{\rm th}(Q^2)$  is given by the OPE expansion (\ref{DOPE}), and the (holomorphic) function $G$ is the following integral of $g$:
\be
G(Q^2)= \int_{-\sm}^{Q^2} d Q^{'2} g(Q^{'2}),
\label{GQ2}
\ee
which is independent of the form of path from $-\sm$ to $Q^2$ in the $Q^{'2}$-complex plane because $g(Q^{'2})$ is holomorphic.

\begingroup\color{black}
The quark vector and axial vector current polarisation functions $\Pi_J^{(k)}(Q^2)$ and the Adler function ${\cal D}(Q^2)$ are quantities which are holomorphic (analytic) functions of $Q^2$ in the complex $Q^2$-plane with the exception of the negative semiaxis, i.e., they are spacelike quantities. On the other hand, the spectral function (\ref{om1}) and the sum rules (\ref{sr}) are timelike observables, being functions of the squared energy $\sigma>0$ ($=-Q^2$) or $\sm \equiv \sigma_{\rm max} >0$. Several other timelike quantities exist that have the form of integrals of ${\cal D}(Q^2)$ and are of phenomenological interest \cite{Nesterenko:2016pmx}, such as: (a) $e^+ e^- \to$ hadrons production ratio $R(s)$ \cite{AKR,ANR} (with any $s>0$); (b) the leading order hadronic vacuum polarisation contribution to the muon anomalous magnetic moment $(g_{\mu}-2)$ \cite{amurev,amuZoltan}, where the dominant squared timelike momenta are $s \sim m^2_{\mu}$ ($\sim 0.01 \ {\rm GeV}^2$) \cite{NestJPG42,amuO}.
\endgroup

\section{Renormalon-motivated extension of the Adler function}
\label{sec:renmod}

Here we summarise the main results of the renormalon-motivated extension of the Adler function. We refer for details to \cite{renmod,EPJ21}. The expansion of the $D=0$ part of the Adler function in powers of $a \equiv \alpha_s/\pi$ has the form (\ref{dpt}), where the first four expansion coefficients ($d_0=1; d_1; d_2; d_3$) are exactly known \cite{d1,d2,BCK}. If we reorganise this expansion into the expansion in terms of the (related) logarithmic derivatives
\be
\ta_{n+1}(Q^{'2}) \equiv \frac{(-1)^n}{n! \beta_0^n} \left( \frac{d}{d \ln Q^{'2}} \right)^n a(Q^{'2}) \qquad (n=0,1,2,\ldots),
\label{tan} \ee
we obtain
\be
d(Q^2)_{D=0, {\rm lpt}}= {\td}_0 a(\kappa Q^2) + {\td}_1(\kappa) \; {\ta}_2(\kappa Q^2) + \ldots + {\td}_n(\kappa) \; {\ta}_{n+1}(\kappa Q^2) + \ldots.
\label{dlpt} \ee
By the use of the $\MSbar$ scheme RGE (which is known up to five-loops \cite{5lMSbarbeta})
\be
\frac{d a(\kappa Q^2)}{d \ln \kappa} = - \beta_0 a(\kappa Q^2)^2 -\beta_1 a(\kappa Q^2)^3 - \sum_{j=2}^4 {\beta}_j a(\kappa Q^2)^{j+2},
\label{RGE}
\ee
the logarithmic derivatives $\ta_{n}$ and the powers $a^k$ can be related
\bes
\label{tananantan}
\bea
\ta_{n+1}(Q^{'2}) &=& a(Q^{'2})^{n+1} + \sum_{m \geq 1} k_m(n+1) \; a(Q^{'2})^{n+1+m},
\label{tanan} \\
a(Q^{'2})^{n+1} &=&  \ta_{n+1}(Q^{'2}) + \sum_{m \geq 1} \tk_m(n+1) \; \ta_{n+1+m}(Q^{'2}),
\label{antan} \eea \ees
and the expansion coefficients $\td_n$ and $d_k$ are related analogously
\bes
\label{tdndndntdn}
\bea
\td_n(\kappa) &=& d_n(\kappa) + \sum_{s=1}^{n-1} \tk_s(n+1-s) \; d_{n-s}(\kappa) \qquad (\td_0=d_0=1),
\label{tdndk} \\
d_n(\kappa) &=& \td_n(\kappa) + \sum_{s=1}^{n-1} k_s(n+1-s) \; \td_{n-s}(\kappa) \quad
(n=0,1,2, \ldots),
\label{dntdk} \eea \ees
The coefficients $\tk_s(n+1-s)$ and $k_s(n+1-s)$ are specific $\kappa$-independent combinations of the $\beta$-function coefficients $c_j = \beta_j/\beta_0$, cf.~\cite{renmod}. The new expansion coefficients $\td_n$ allow us to construct a Borel transform ${\cal B}[\td](u)$ related to the original Borel transform ${\cal B}[d](u)$ of the Adler function (\ref{Bdexp})
\be
{\cal B}[\td](u; \kappa) \equiv {\td}_0 + \frac{{\td}_1(\kappa)}{1! \beta_0} u + \ldots + \frac{{\td}_n(\kappa)}{n! \beta_0^n} u^n + \ldots,
\label{Btdexp} \ee
which contains all the information about the Adler function coefficients $\td_n$ (and thus $d_n$) but, in contrast to  ${\cal B}[d](u)$, has the simple one-loop type renormalisation scale dependence
\be
\frac{d}{d \ln \kappa} {\td}_n(\kappa) = n \beta_0 {\td}_{n-1}(\kappa) \quad \Rightarrow \quad {\cal B}[\td](u; \kappa) = \kappa^u {\cal B}[\td](u).
\label{tdnkap} \ee
\textcolor{black}{As a consequence, the renormalon structure of ${\cal B}[\td](u)$ has no explicit effects coming from the beyond one-loop RGE running of the coupling, i.e., no terms $\propto \beta_1$ in the power indices of the singularities [cf.~Eq.~(\ref{RGE})]. We adopt the approximation that the renormalon structure of  ${\cal B}[\td](u)$ has the form} as obtained in the LB approximation \cite{LB1,LB2}: ${\cal B}[\td](u) \sim 1/(2-u)$, $1/(3-u)^2$, $1/(3-u), \ldots$ near the IR renormalon locations\footnote{
Near the leading IR renormalon location $u=2$, it is reasonable to include a subleading singularity $\sim \ln(1 - u/2)$ in  ${\cal B}[\td](u)$, which corresponds in the Borel transform of the Adler function to the subleading singularity ${\cal B}[d](u) \sim 1/(2-u)^{\tg_2-1}$, cf.~\cite{renmod}.}
  $u=2,3,\ldots$; and ${\cal B}[\td](u) \sim 1/(1+u)^2$, $1/(1+u)$, $1/(2+u)^2$, $1/(2+u), \ldots$ near the UV renormalon locations $u=-1, -2, \ldots$ As shown in \cite{renmod}, this then implies that the Borel transform ${\cal B}[d](u)$ of the Adler function behaves near these renormalon locations as the theory suggests: ${\cal B}[d](u) \sim 1/(2-u)^{\tg_2}$, $1/(3 - u)^{\tg_3+1}$, $1/(3 - u)^{\tg_3}, \ldots$ near $u=2,3,\ldots$ where $\tg_p = 1 + p \beta_1/\beta_0^2$; and  ${\cal B}[d](u) \sim 1/(1+u)^{\bg_1+1}$, $1/(1+u)^{\bg_1}$, $1/(2+u)^{\bg_2+1}$, $1/(2+u)^{\bg_2}, \ldots$ near $u=-1,-2,\ldots$ where $\bg_p =  1 - p \beta_1/\beta_0^2$. 

\textcolor{black}{We point out that the general structure of the IR singularity at $u=p=D/2$ ($p=2,3,\ldots$) is of the form $1/(p-u)^{\cal K}$ where ${\cal K} =1 + p   \beta_1/\beta_0^2 - \gamma_{O_D}^{(1)}/\beta_0$ (using the notation of \cite{renmod}) where $\gamma_{O_D}^{(1)}$ is the effective leading-order anomalous dimension of the corresponding dimension-$D$ OPE operator (which, in this convention, includes the leading-order effects of the corresponding Wilson coefficient). For the case $D=4$ we have $\gamma_{O_4}^{(1)}/\beta_0 =0$ in the LB approximation and exactly (beyond-LB). However, for $D=6$ ($p=3$), this quantity is $\gamma_{O_6}^{(1)}/\beta_0 =-1$ or $0$ only in the LB approximation, but becomes a set of noninteger numbers in the exact case (beyond-LB) \cite{Boito:2015joa}, namely $\gamma_{O_6}^{(1)}/\beta_0 \approx 0.197;$ $0.247; \ldots$, corresponding to a set of nine operators with dimension $D=6$ (cf.~also \cite{JK,LSCh,ACh}). The corresponding $D=6$ OPE terms of the Adler function would then be nine terms $\sim a(Q^2)^{\gamma_{O_6}^{(1)}/\beta_0}/(Q^2)^3$, instead of the two $D=6 $ terms appearing in Eq.~(\ref{DOPE}). We will not pursue this extension in this work, but wish to point out that this indicates the challenging way how one may extend the present renormalon-motivated model to structures which contain improved beyond-LB effects in the IR renormalons $u=p \geq 3$ and the corresponding OPE terms of dimension $D= 2 p \geq 6$.}

Thus the ansatz we make for ${\cal B}[\td](u)$ (with $\kappa=1$) in the $\MSbar$ scheme includes the singularities at the locations $u=2, 3$ and $u=-1$ \cite{renmod}
\bea
{\cal B}[\td](u) & = & \exp \left( \tK u \right) \pi {\Big \{}
\td_{2,1}^{\rm IR} \left[ \frac{1}{(2-u)} + \tal (-1) \ln \left( 1 - \frac{u}{2} \right) \right] + \frac{ \td_{3,2}^{\rm IR} }{(3 - u)^2} + \frac{ \td_{3,1}^{\rm IR} }{(3 - u)} + \frac{ \td_{1,2}^{\rm UV} }{(1 + u)^2} {\Big \}},
\label{Btd5P}
\eea
where the value of the parameter $\tal$ in the $\MSbar$ scheme is fixed, $\tal =-0.255$ \cite{renmod}, \textcolor{black}{by the knowledge of the subleading coefficient ${\hat c}^{(D=4)}_1$ coming from the $D=4$ subleading-order anomalous dimension and Wilson coefficient \cite{CPS,ren}.} The other five parameters ($\tK$ and the residues $\td_{2,1}^{\rm IR}, \td_{3,2}^{\rm IR}, \td_{3,1}^{\rm IR}, \td_{1,2}^{\rm UV}$) are determined by the knowledge of the first five coefficients $d_n$ (and thus $\td_n$), $n=0,1,2,3,4$. However, for $n=4$, the coefficient $d_4$ of the Adler function is not yet exactly known. There are, however, estimates for the value of this coefficient. In Ref.~\cite{renmod}, a similar ansatz as (\ref{Btd5P}) was used in the lattice-related MiniMOM renormalisation scheme, not containing $\td_{3,1}^{\rm IR}/(3-u)$ term, and the resulting estimate $d_4 =338.19$ (in $\MSbar$ scheme, with $\kappa=1$) was extracted upon reexpansion. On the other hand, the effective charge (ECH) method \cite{ECH} gives an estimate $d_4=275$ \cite{KatStar,BCK}; a similar estimate $d_4=277 \pm 51$ was obtained in \cite{Boitoetal} based on Pad\'e approximants; \textcolor{black}{$d_4=283$ was obtained in \cite{BJ} by extrapolating the approximately geometric series behaviour of FOPT of $r_{\tau}^{(D=0)}$ [$=a^{(2,1)}(m_{\tau}^2)$].}  We will use in our renormalon-motivated model (\ref{Btd5P}) the value $d_4 =275.$ as the central value, and will include the value $d_4=338.19$ via variation
\begingroup \color{black}
\be
d_4 = 275. \pm 63.19
\label{d4est} \ee
\endgroup
In Table \ref{tabrenmod1} we present the resulting parameters $\tK$ and $\td_{i,j}^{\rm IR}$ ($i=2,3$; $j=1,2$) and $\td_{1,2}^{\rm UV}$ for the central and the border values of $d_4$ Eq.~(\ref{d4est}). In Table \ref{tabrenmod2}, on the other hand, we present the values of the first 11 coefficients $\td_n$ and $d_n$ of the Adler function, for the three mentioned values of $d_4$. 
\begin{table}
  \caption{The values of $\tK$ and of the renormalon residues $\td_{i,j}^{\rm X}$ (X=IR,UV) for the five-parameter ansatz (\ref{Btd5P}) in the $\MSbar$ scheme, when $d_4$ is taken according to Eq.~(\ref{d4est}).}
\label{tabrenmod1}
\begin{ruledtabular}
\begin{tabular}{r|rrrrr}
  $d_4$ & $\tK$ & $\td_{2,1}^{\rm IR}$ & $\td_{3,2}^{\rm IR}$ &  $\td_{3,1}^{\rm IR}$ & $\td_{1,2}^{\rm UV}$ 
\\
\hline
275.                & 0.16010 & 0.661852 & 2.04546 & -0.68316 & -0.0121699
\\
275.  - 63.19       & -0.33879 & 0.986155 & 6.75278 & -2.74029 & -0.011647
\\
275.  + 63.19       & 0.5190 & 1.10826 & -0.481538 & -0.511642 & -0.0117704
\end{tabular}
\end{ruledtabular}
\end{table}
\begin{table}
  \caption{The $\MSbar$ coefficients $\td_n$ and $d_n$ (with $\kappa=1$) ($n \leq 10$) for the three cases of $d_4$ Eq.~(\ref{d4est}).}
\label{tabrenmod2}
\begin{ruledtabular}
\begin{tabular}{r|rr|rr|rr}
 $n$ & $d_4=275.$: $\td_n$ & $d_n$ & $d_4=211.81$: $\td_n$ & $d_n$ & $d_4=338.19$: $\td_n$ & $d_n$ 
\\
\hline
0       &  1       & 1           &  1       & 1  &  1       & 1   \\
1       &  1.63982 & 1.63982    &  1.63982 & 1.63982  &  1.63982 & 1.63982 \\
2       &  3.45578 & 6.37101    &  3.45578 & 6.37101  &  3.45578 & 6.37101\\
3       &  26.3849 & 49.0757   &  26.3849 & 49.0757 &  26.3849 & 49.0757\\
4       & -25.4181 & 275. &  -88.6087 & 211.81 & 37.7719 & 338.19 \\
5       &  1812.35 & 3159.46 & 2307.09 & 2933.36 & 1732.04 & 3799.99 \\
6       & -19073.5 & 16136.2 & -30145.1 & 7061.98 & -9949.19 & 29672.9 \\
7       &  411373. & 340795. & 591695. & 332609. & 322129. & 465315. \\
8       & $-7.72963 \times 10^{6}$ & 378157. & $-1.16707 \times 10^{7}$ & $-1.42196 \times 10^{6}$ & $-5.1117 \times 10^{6}$ & $3.21051 \times 10^{6}$ \\
9       & $1.79184 \times 10^{8}$ & $6.99442 \times 10^{7}$ & $2.67043 \times 10^{8}$ & $8.1617 \times 10^{7}$ &  $1.28702 \times 10^{8}$ & $8.8993 \times 10^{7}$ \\
10      & $-4.36871 \times 10^{9}$ & $-5.83093 \times 10^{8}$ & $-6.59462 \times 10^{9}$ & $-1.19984 \times 10^{9}$ & $-3.00623 \times 10^{9}$ & $1.7999 \times 10^{8}$
\end{tabular}
\end{ruledtabular}
\end{table}
\begingroup\color{black}
The expression (\ref{Btd5P}) then implies, as explained above, for the Borel transform of the Adler function ${\cal B}[d](u)$ the following expression near the singularities (for $\kappa=1$):
\bea
\lefteqn{
  \frac{1}{\pi} {\cal B}[d](u) \equiv  \frac{1}{\pi} \left[ d_0 + \frac{d_1(\kappa)}{1! \beta_0} u + \ldots + \frac{d_n(\kappa)}{n! \beta_0^n} u^n + \ldots \right]
  }
\nonumber\\
& = &
{\bigg \{} \frac{d_{2,1}^{\rm IR}(\tkap)}{(2 - u)^{\tg_2}}
\left[ 1 -1.70601 \; (2-u) -0.934455  \; (2-u)^2 \right]  + {\cal O} \left( (2-u)^{-\tg_2+3} \right)
{\bigg \}}
\nonumber\\ &&
{\bigg \{}
+ \frac{d_{3,2}^{\rm IR}(\tkap)}{(3 - u)^{\tg_3+1}} \left[ 1 -2.0284  \; (3-u) +2.05317  \; (3-u)^2 \right]
\nonumber\\ &&
+ \frac{d_{3,1}^{\rm IR}(\tkap)}{(3 - u)^{\tg_3}} \left[ 1 -1.54726  \; (3-u)  \right] + {\cal O} \left( (3-u)^{-\tg_3+2} \right)
{\bigg \}}
\nonumber\\ &&
{\bigg \{}
+ \frac{d_{1,2}^{\rm UV}(\tkap)}{(1+ u)^{\bg_1+1}} \left[ 1 + 16.3722  \; (1+u) + 116.764  \; (1 + u)^2  \right] + {\cal O} \left( (1 + u)^{-\bg_1 + 2} \right)
{\bigg \}}.
\nonumber\\
\label{Bd5P}
\eea
where $\tg_p = 1 + p \beta_1/\beta_0^2$; $\bg_p=1 - p \beta_1/\beta_0^2$, and the residues in the expression (\ref{Bd5P}) are
\bes
\bea
d_{2,1}^{\rm IR} & = & 1.64049,
\label{d21IR} \\
d_{3,2}^{\rm IR}&=&98.2065, \qquad d_{3,1}^{\rm IR}=-14.7528, 
\label{d3jIR} \\
  d_{1,2}^{\rm UV} & = & -0.0109501.
\label{d12UV}
\eea
\ees
For comparison with the expression (\ref{Bd5P}), in the renormalon model of Refs.~\cite{BJ,BJ2}, on the other hand, the Borel transform  ${\cal B}[d](u)$ is constructed by a direct ansatz [i.e., not via ${\cal B}[\td](u)$ Eq.~(\ref{Btd5P})], and consists of the terms $\sim 1/(2 - u)^{\tg_2}$,  $\sim 1/(3 - u)^{\tg_3}$, $\sim 1/(1+ u)^{\bg_1+1}$ and a linear polynomial of $u$. That model was later referred to and compared with Pad\'e-related approaches in Ref.~\cite{Boitoetal}.
\endgroup

\section{Specific sum rules used}
\label{sec:SR}

Once having the expansion coefficients of $d(Q^2)_{D=0}$, we can apply various sum rules, i.e., various weight functions $g(Q^2)$ [cf.~Eqs.~(\ref{sr})-(\ref{GQ2} We will primarily consider the double-pinched Borel-Laplace transforms $B(M^2)$ where $M^2$ is a complex squared energy parameter\footnote{Double-pinched means that the weight functions $g(Q^2)$ have double zero at the Minkowskian point $Q^2=-\sm$, which effectively suppresses the duality violation effects, cf.~e.g.~Ref.~\cite{Pich}.}
\bes
\label{BL}
\bea
g_{M^2}(Q^2) &=&  \left( 1 + \frac{Q^2}{\sm} \right)^2  \frac{1}{M^2} \exp \left( \frac{Q^2}{M^2} \right) \qquad \Rightarrow
\label{gM2} \\
G_{M^2}(Q^2) & = &  \left\lbrace  \left[ \left( 1 + \frac{Q^2}{\sigma_m} \right)^2 - 2\frac{M^2}{\sigma_m} \left(1 + \frac{Q^2}{\sigma_m}\right) + 2 \left(\frac{M^2}{\sigma_m} \right)^2 \right] \exp\left(\frac{Q^2}{M^2}\right) - 2 \left( \frac{M^2}{\sigma_m} \right)^2 \exp \left( -\frac{\sigma_m}{M^2}\right) \right\rbrace,
\label{GM2}
\eea \ees
The general sum rule Eq.~(\ref{sr}) has on the left-hand side the experimental value, and on the right-hand side the theoretical value. Using the (double-pinched) Borel-Laplace weight function $g(Q^2)$ Eq.~(\ref{gM2}), the left-hand side of the sum rule is written as
\be
B_{\rm exp}(M^2;\sm) =  \int_0^{\sm} d \sg \; g_{M^2}(-\sg) \omega_{\rm exp}(\sg)
= \frac{1}{M^2} \int_0^{\sm} d \sg \; \left( 1 - \frac{\sg}{\sm} \right)^2  \exp \left( -\frac{\sg}{M^2} \right) \omega_{\rm exp}(\sg),
\label{Bexp}
\ee
and the right-hand side as
\bea
B_{\rm th}(M^2; \sm) &=&   \frac{1}{2 \pi} \int_{-\pi}^{+\pi} d \phi \;
  G_{M^2} \left(\sm e^{i \phi} \right) {\cal D}_{\rm th} \left( \sm e^{i \phi} \right)
  \nonumber\\ &&
\!\!\!\!\!\!\!\!\!\!\!\!\!\!\!\!\!\!\!\!\!\!\!\!\!\!
  =  \left[ \left( 1 - 2 \frac{M^2}{\sm} \right) + 2 \left( \frac{M^2}{\sm} \right)^2 \left(1 - \exp \left( - \frac{\sm}{M^2} \right) \right) \right]  
  \nonumber\\ &&
\!\!\!\!\!\!\!\!\!\!\!\!\!\!\!\!\!\!\!\!\!\!\!\!\!\!\!  
+ \frac{1}{2 \pi}  \int_{-\pi}^{+\pi} d \phi \left\{ \left[ \left(1 + e^{i \phi} \right)^2 - 2 \frac{M^2}{\sm}  \left(1 + e^{i \phi} \right) +  2 \left( \frac{M^2}{\sm} \right)^2 \right] \exp \left( \frac{\sm}{M^2} e^{i \phi} \right) -  2 \left( \frac{M^2}{\sm} \right)^2 \exp \left( - \frac{\sm}{M^2} \right) \right\} d \left( \sm e^{i \phi} \right)_{D=0}
  \nonumber\\ &&
\!\!\!\!\!  + B_{\rm th}(M^2; \sm)_{D=4} + \sum_{k \geq 3} B_{\rm th}(M^2; \sm)_{D=2 k}.
\label{Bth}
\eea
The last terms here are the contributions of the dimension $D=2 k$ condensates of the OPE of the Adler function (\ref{DOPE})
\bes
\label{BthD}
\bea
B_{\rm th}(M^2; \sm)_{D=4} & = &  \frac{2 \pi^2 \langle O_4 \rangle}{ (M^2)^2 } \left( 1 + 2 \frac{M^2}{\sm}  \right),
\label{BD4} \\
B_{\rm th}(M^2; \sm)_{D=2 k} &=& \frac{2 \pi^2}{(k-1)! (M^2)^k}\left( \langle O^{(1)}_{2 k} \rangle + \frac{\langle O^{(2)}_{2 k} \rangle}{a(\sm)} \right) 
\left[ 1 + 2 (k-1) \frac{M^2}{\sm} + (k-1) (k-2) \left(\frac{M^2}{\sm}\right)^2 \right]
\nonumber\\
&& + \frac{2 \pi^2 k}{\sm^k} \langle O^{(2)}_{2 k} \rangle \beta_0 {\Bigg \{}
\left[ \left(1 - 2 \frac{M^2}{\sm} + 2 \left( \frac{M^2}{\sm} \right)^2 \right) J_k \left( \frac{\sm}{M^2} \right) + 2 \left( 1 -  \frac{M^2}{\sm}  \right) J_{k-1}\left( \frac{\sm}{M^2} \right) + J_{k-2}\left( \frac{\sm}{M^2} \right) \right]
\nonumber\\ &&
+ 2 \left( \frac{M^2}{\sm} \right)^2 \exp \left[ -  \frac{\sm}{M^2} \right] \frac{(-1)^k}{k} {\Bigg \}}, 
\label{Bth2k}
\eea \ees
where $k=3,4, \ldots$. The coupling $a(Q^2)$ at the OPE term $\langle {\cal O}^{(2)}_{2 k} \rangle/a(Q^2)$ will be taken to run according to one-loop RGE along the contour  $|Q^2|=\sm$
\be
\frac{\langle {\cal O}^{(2)}_{2 k} \rangle}{a(\sm e^{i \phi})} =  \langle {\cal O}^{(2)}_{2 k} \rangle \left( \frac{1}{a(\sm)} + i \beta_0 \phi \right) ,
\label{O2arun} \ee
and\footnote{If this $1/a(Q^2)$ is run according to 5-loop RGE (instead of one-loop RGE) along the contour, the numerical results do not change significantly.}
$J_s(A)$ are the integrals
\be
J_s(A) \equiv \frac{1}{2 \pi} \int_{-\pi}^{+\pi} d \phi \; \exp \left[ A e^{i \phi} \right]e^{-i s \phi} i \phi.
\label{Js} \ee
The expressions for $J_s(A)$ in terms of sums, for any complex $A$ and any positive integer $s$, are given in the Appendix. Further, explicit closed expressions are given there for $J_s(A)$ with $s=1,2,3$ (which are relevant for the $D= 6$ contribution).

On the other hand, one can use FESRs with (double-pinched) momenta $a^{(2,n)}$ which are associated with the following weight functions $g^{(2,n)}$ ($n=0,1,\ldots$):
\bes
\label{FESR}
\bea
g^{(2,n)}(Q^2) &=& \left( \frac{n + 3}{n + 1} \right)\frac{1}{\sigma_m} \left( 1 + \frac{ Q^2}{\sigma_m} \right)^2 \sum_{k = 0}^{n} (k + 1)(-1)^k\left(\frac{Q^2}{\sigma_m} \right)^k \nonumber \\
 &=& \left( \frac{n + 3}{n + 1} \right)\frac{1}{\sigma_m} \left[ 1 - (n + 2) \left( - \frac{Q^2}{\sigma_m}\right)^{n + 1}  + (n + 1) \left(-\frac{Q^2}{\sigma_m} \right)^{n + 2}  \right] \; \Rightarrow 
\label{g2n}
\\
 G^{(2,n)}(Q^2) &=& \left( \frac{n + 3}{n + 1} \right)\frac{Q^2}{\sigma_m} \left[ 1 - \left(- \frac{Q^2}{\sigma_m} \right)^{n + 1}   \right] + \left[1 -  \left(-\frac{Q^2}{\sigma_m} \right)^{n + 3} \right].
\label{G2n}
\eea \ees
The experimental and the theoretical parts of these FESR moments are then (we subtract unity for convenience)
\bes
\label{a2n}
\bea
a_{\rm exp}^{(2,n)}(\sigma_m) &=& \int_{0}^{\sigma_m} d\sigma \; g^{(2,n)}(-\sigma) \omega_{\rm exp}(\sigma) - 1
\label{a2nexp} \\
 a_{\rm th}^{(2,n)}(\sigma_m) &=&  \frac{1}{2\pi} \int_{-\pi}^{\pi} d\phi \; G^{(2,n)}(\sigma_m e^{i\phi}) \left[ D_{\rm th}(\sigma_m e^{i\phi}) - 1 \right]
 \label{a2nth} \eea \ees
 We will consider in particular the first two moments $a^{(2,0)}$ and $a^{(2,1)}$, up to $D=6$ terms
 \bes
 \label{a2021}
 \bea
 \sum_{k \geq 2} a^{(2,0)}_{\rm th}(\sm)_{D=2 k} = \frac{12 \pi^2}{\sm^2} \langle O_4 \rangle + \frac{6 \pi^2}{\sm^3} \left\{ \left[ \langle O^{(1)}_6 \rangle + \frac{\langle O^{(2)}_6 \rangle}{a(\sm)} \right] + \frac{11}{6} \beta_0 \langle O^{(2)}_6 \rangle \right\} + {\cal O}\left( \frac{1}{\sm^4} \right),
 \label{a20th} \\
 \sum_{k \geq 2} a^{(2,1)}_{\rm th}(\sm)_{D=2 k} =  \frac{12 \pi^2}{\sm^3} \left\{ - \left[ \langle O^{(1)}_6 \rangle + \frac{\langle O^{(2)}_6 \rangle}{a(\sm)} \right] + \frac{1}{6} \beta_0 \langle O^{(2)}_6 \rangle \right\} + {\cal O}\left( \frac{1}{\sm^4} \right).
 \label{a21th} \eea \ees
The expressions for $a^{(2,n)}_{\rm th}(\sm)_{D=2 k}$ for general integer $n \geq 0$ (and $k \geq 2$) are given in the Appendix. We note that $a^{(2,1)}(\sigma=m^2_{\tau})_{D=0}$ is the $D=0$ part of the canonical QCD (and strangeless and massless) $\tau$-decay ratio $r_{\tau}^{(D=0)}$, cf.~footnote \ref{rtau}.
 
\section{Methods of evaluation of the $D=0$ contribution}
\label{sec:methD0}

We will use four different methods for the evaluation of the truncated $D=0$ contributions to the Sum Rules: Fixed Order Perturbation Theory using powers (FO); Fixed Order Perturbation Theory using logarithmic derivatives (${\widetilde {\rm FO}}$ or tFO); Contour Improved Perturbation Theory (CI); Inverse Borel Transformation with Principal Value (PV).

\begin{enumerate}
\item Fixed Order Perturbation Theory using powers (FO): The truncated power expansion $d(\sm e^{i \phi})^{[N_t]}_{D=0, {\rm pt}}$ [cf.~Eq.~(\ref{dpt})]\footnote{The expression (\ref{dpttr}) has some dependence on the renormalisation scale parameter $\kappa$ due to truncation, $(d/ d \ln \kappa) d(Q^2; \kappa)^{[N_t]}_{D=0, {\rm pt}} \sim a^{N_t+1}$.}
\be
d(\sm e^{i \phi}; \kappa)^{[N_t]}_{D=0, {\rm pt}} = a(\kappa \sm e^{i \phi}) + \sum_{n=1}^{N_t-1} d_n(\kappa) a(\kappa \sm e^{i \phi})^{n+1},
\label{dpttr} \ee
which appears in the contour integrals in the sum rules, Eqs.~(\ref{Bth}) and (\ref{a2nth}), is written as truncated Taylor expansion in powers of $a(\kappa \sm)$ up to (and including) $a(\kappa \sm)^{N_t}$. We point out that the Adler function $d(Q^2)_{D=0}$ is a spacelike quantity (defined for general complex $Q^2$); the sum rules are timelike quantities (defined only for positive quantities $\sigma =\sm >0$), but are written in the FO approach in terms of powers of $a(\kappa \sm)$  where $Q^2 = \kappa \sm > 0$ is a spacelike point in the complex $Q^2$-plane.
\item Fixed Order Perturbation Theory using logarithmic derivatives (${\widetilde {\rm FO}}$): The truncated expansion $d(\sm e^{i \phi})^{[N_t]}_{D=0, {\rm lpt}}$ [cf.~Eq.~(\ref{dlpt})]
\be
d(\sm e^{i \phi}; \kappa)^{[N_t]}_{D=0, {\rm lpt}} = a(\kappa \sm e^{i \phi}) + \sum_{n=1}^{N_t-1} {\td}_n(\kappa) \; {\ta}_{n+1}(\kappa \sm e^{i \phi})
\label{dlpttr} \ee
in the countour integrals is written as truncated Taylor expansion in logarithmic derivatives ${\ta}_{k+1}(\kappa \sm)$ up to (and including) ${\ta}_{N_t}(\kappa \sm)$.
\item Contour Improved Perturbation Theory (CI): The truncated power expansion (\ref{dpttr}) in the contour integrals is kept as it is, $a(\kappa \sm e^{i \phi})$ is the (five-loop) RGE-running coupling.
\item Inverse Borel Transformation with Principal Value (PV):
  The expression for the $D=0$ part of the Adler function in the contour integrals is written as
\be
{\bigg (} d(\sm e^{i \phi})_{D=0} {\bigg )}^{({\rm PV}, [N_t])} = \frac{1}{\beta_0} \frac{1}{2} \left( \int_{{\cal C}_{+}} + \int_{{\cal C}_{-}} \right) d u \exp \left[ - \frac{u}{\beta_0 a(\kappa \sm e^{i \phi})} \right] {\cal B}[d](u; \kappa)_{\rm sing} + \delta d(\sm e^{i \phi}; \kappa)^{[N_t]}_{D=0},
\label{PV2}
\ee 
where ${\cal B}[d](u; \kappa)_{\rm sing}$ is the truncated singular part of the Borel transform of $d(Q^2)_{D=0}$; \textcolor{black}{in the case $\kappa=1$ it is given by Eq.~(\ref{Bd5P}) without the terms indicated there as ${\cal O}(\ldots)$.} The arithmetic average over the integration paths ${\cal C}_{\pm}$ gives the Principal Value. The expression $\delta d(\sm e^{i \phi}; \kappa)^{[N_t]}_{D=0}$ is the truncated series in powers of $a(\sm e^{i \phi})$ which completes the power terms corresponding to the Inverse Borel Transform of the singular part, \textcolor{black}{i.e., it accounts up to $a^{N_t}$ for the terms ${\cal O}((2-u)^{-\tg_2+3})$, ${\cal O}((3-u)^{-\tg_3+2})$ and ${\cal O}((1+u)^{-\bg_1+2})$ not included in ${\cal B}[d](u; \kappa)_{\rm sing}$ [cf.~Eq.~(\ref{Bd5P})]. The coefficients $\delta d_n$ of the series  $\delta d(\sm e^{i \phi}; \kappa)^{[N_t]}_{D=0}$ are thus practically free of renormalon growth when $n$ increases.} We refer for additional explanation to \cite{EPJ21} (Sec.~IV.B there).
\end{enumerate}

\section{Results of fitting the Borel-Laplace sum rules}
\label{sec:res}

In this Section we first fit the theoretical double-pinched Borel-Laplace sum rules, cf.~Eqs.~(\ref{sr}) and (\ref{Bexp})-(\ref{BthD}), to the ALEPH experimental data (V+A channel) as explained in Sec.~\ref{sec:SRgen}. The theoretical Borel-Laplace sum rules $B_{\rm th}(M^2;\sm)$ are evaluated with various evaluation methods and various truncation indices $N_t$ in the $D=0$ contribution, cf.~Sec.~\ref{sec:methD0} for explanations. Subsequently, the resulting predictions for the (double-pinched) FESR momenta $a^{(2,0)}(\sm)$ and $a^{(2,1)}(\sm)$ [cf.~Eqs.~(\ref{a2n})-(\ref{a2021})] are compared with the experimental data, for various truncation indices $N_t$, and the optimal $N_t$ is fixed where the relative stability of these FESRs under variation of $N_t$ is reached. We point out that in the analysis, the higher order contributions of the $D=0$ Adler function contributions are generated (estimated) by the renormalon-motivated ansatz mentioned in Sec.~\ref{sec:renmod}. Further, throughout the analysis, the OPE expansion (\ref{DOPE}) of the Adler function is performed up to $D (\equiv 2 k) =6$ terms. Going beyond $D=6$ terms is not well motivated in the present analysis, as the $D=0$ Adler function is generated by a renormalon-motivated ansatz which includes IR renormalons up to $u=+3$ and not beyond. This means that an assumption is made that the higher IR renormalons ($u=+4$, etc.) give suppressed contributions to the $D=0$ Adler function; such an assumption would then also suggest that the $D \geq 8$ OPE contributions to the Adler function are in general suppressed.

In practice, the double-pinched Borel Laplace sum rule is applied to the real parts
\be
{\rm Re} B_{\rm exp}(M^2;\sm) = {\rm Re} B_{\rm th}(M^2;\sm),
\label{BLsr} \ee
where for the Borel-Laplace scale parameters $M^2$ we take $M^2 = |M^2| \exp(i \Psi)$, where $0 \leq \Psi < \pi/2$. Specifically, we take $0.9 \ {\rm GeV}^2 \leq |M^2| \leq 1.5 \ {\rm GeV}^2$, and $\Psi = 0, \pi/6, \pi/4$. The choices of these values were motivated in \cite{EPJ21}. In practice, we minimised the difference between the two quantities (\ref{BLsr}) by minimising the following sum of squares:
\be
\chi^2 = \sum_{\alpha=0}^n \left( \frac{ {\rm Re} B_{\rm th}(M^2_{\alpha};\sm) - {\rm Re} B_{\rm exp}(M^2_{\alpha};\sm) }{\delta_B(M^2_{\alpha})} \right)^2 ,
\label{xi2} \ee
where $M_{\alpha}^2$ is a dense set of points along the chosen rays with $\Psi=0, \pi/6, \pi/4$  and $0.9 \ {\rm GeV}^2 \leq |M|^2 \leq 1.5 \ {\rm GeV}^2$. Specifically, we chose 11 equidistant points along each of the three rays.\footnote{The sum (\ref{xi2}) thus contains 33 terms; but the fit results remain practically unchanged when the number of points is increased.} In the sum (\ref{xi2}), the quantities $\delta_B(M^2_{\alpha})$ are the experimental standard deviations of ${\rm Re} B_{\rm exp}(M^2_{\alpha};\sm)$, cf.~\cite{EPJ21} for more explanation.

\begin{figure}[htb] 
\centering\includegraphics[width=110mm]{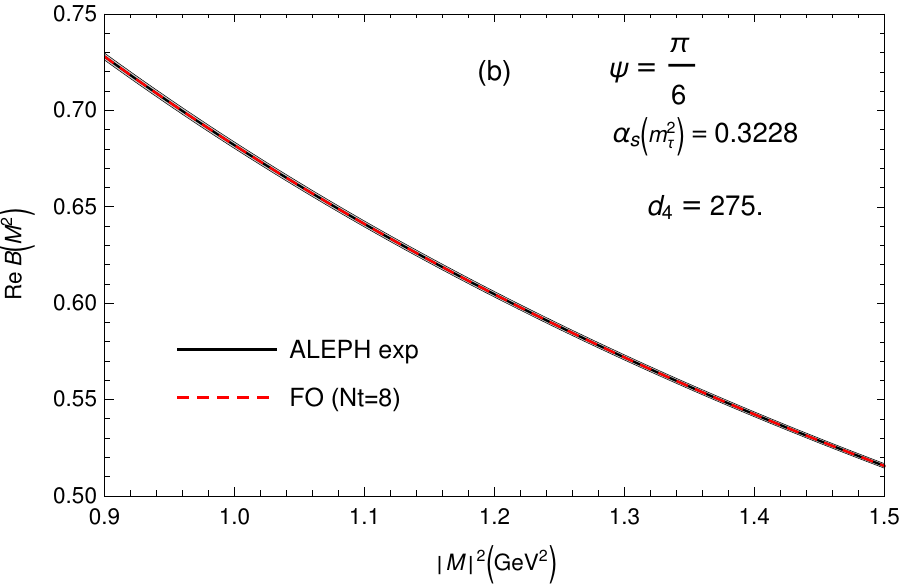}
\caption{(coloured online) The values of ${\rm Re} B(M^2;\sigma_{\rm m})$ along the ray $M^2=|M^2| \exp( i \Psi)$ with $\Psi=\pi/6$. The narrow grey band are the experimental predictions. The red dashed line inside the band is the result of the FOPT global fit with truncation index $N_t=8$. For $\Psi=0$ and $\pi/4$, similarly good fitted curves are obtained.}
\label{fig:FigPsiPi6}
\end{figure}

The expressions ${\rm Re} B_{\rm th}(M^2_{\alpha};\sm)$ depend on four different parameters appearing in the OPE (\ref{DOPE}) of the Adler function with $D \leq 6$: $\alpha_s$,  $\langle O_4 \rangle$, $\langle O_6^{(1)} \rangle$ and  $\langle O_6^{(2)} \rangle$. The minimisation of $\chi^2$ Eq.~(\ref{xi2}) is performed by varying these four parameters simultaneously. In most cases the fits are very good, namely $\chi^2 \lesssim 10^{-3}$, , cf.~Fig.~\ref{fig:FigPsiPi6}.

In Table \ref{tabBL} we present the results of this analysis.\footnote{
From the above values of $\langle O_4 \rangle_{V+A}$, the corresponding values for the gluon condensate are obtained by using the relation $\langle a GG \rangle = 6  \langle O_4 \rangle + 6 f^2_{\pi} m^2_{\pi}$, where $6 f^2_{\pi} m^2_{\pi} \approx 0.00199 \ {\rm GeV}^4$.}
\begin{table}
  \caption{The results for $\alpha_s(m_{\tau}^2)$ and the three condensates (of the full V+A channel) as obtained by the Borel-Laplace sum rule. Included are the optimal truncation numbers ($N_t$) and the values of the fit quality $\chi^2$ [cf.~the text and Eq.~(\ref{xi2})]. Note that $\langle O^{(2)}_6 \rangle$ is given in units of $10^{-4} \ {\rm GeV}^6$ because it is in general smaller than $\langle O^{(1)}_6 \rangle$ by one order of magnitude.}
 \label{tabBL}
\begin{ruledtabular}
\begin{tabular}{r|rrrr|r|r}
  method & $\alpha_s(m_{\tau}^2)$ &  $\langle O_4 \rangle$ ($10^{-3} \ {\rm GeV}^4$) & $\langle O^{(1)}_6 \rangle$ ($10^{-3} \ {\rm GeV}^6$)  & $\langle O^{(2)}_6 \rangle$ ($10^{-4} \ {\rm GeV}^6$) & $N_t$ & $\chi^2$ \\
\hline
FOPT                     &  $0.3228^{+0.0134}_{-0.0121}$      & $-5.3^{+4.8}_{-5.2}$   &  $-2.8^{+1.5}_{-0.1}$  &  $+7.1^{+2.3}_{-4.4}$ & 8 & $1.3 \times 10^{-3}$ \\
${\widetilde{\rm FOPT}}$ &  $0.3209^{+0.0207}_{-0.0298}$      & $-4.6^{+6.0}_{-14.5}$   &  $-3.2^{+3.7}_{-0.3}$  &  $+7.7^{+6.1}_{-8.1}$ & 5 & $2.9 \times 10^{-3}$ \\
CIPT                     &   $0.3488^{+0.0142}_{-0.0028}$ &  $-3.6^{+1.5}_{-3.8}$ & $-3.2^{+2.1}_{-6.8}$ & $+5.8^{+10.1}_{-2.8}$ & 4 &  $0.5 \times 10^{-3}$ \\
PV      &  $0.3269^{+0.0265}_{-0.0092}$ & $-4.6^{+2.4}_{-7.0}$ & $-6.7^{+5.3}_{-19.0}$ & $+8.9^{+26.4}_{-6.0}$ & 8 &  $3.1 \times 10^{-3}$
\end{tabular}
\end{ruledtabular}
\end{table}
The uncertainties in the Table were obtained by combining various theoretical uncertainties and the experimental uncertainty, as will be explained below in more detail for the case of the parameter $\alpha_s(m_{\tau}^2)$.

\begin{figure}[htb] 
\begin{minipage}[b]{.49\linewidth}
  \centering\includegraphics[width=85mm]{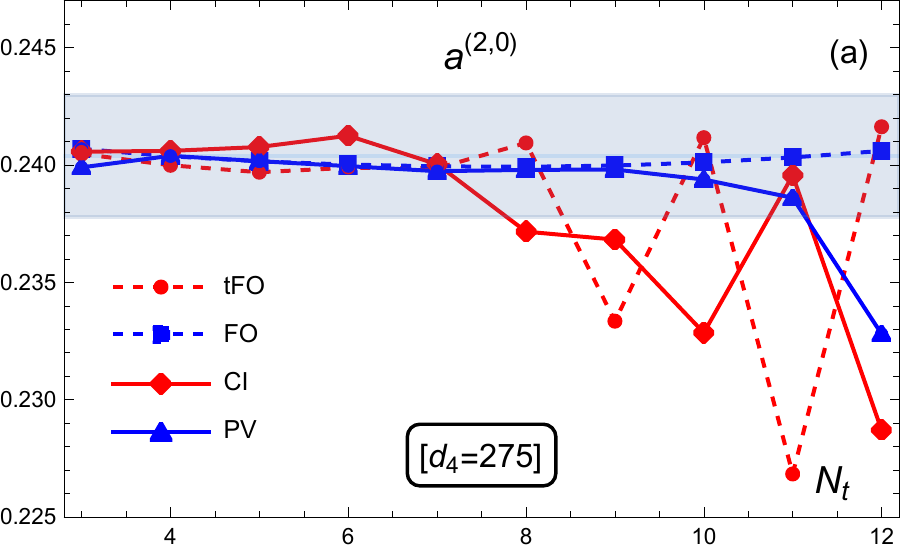}
  \end{minipage} 
\begin{minipage}[b]{.49\linewidth}
  \centering\includegraphics[width=85mm]{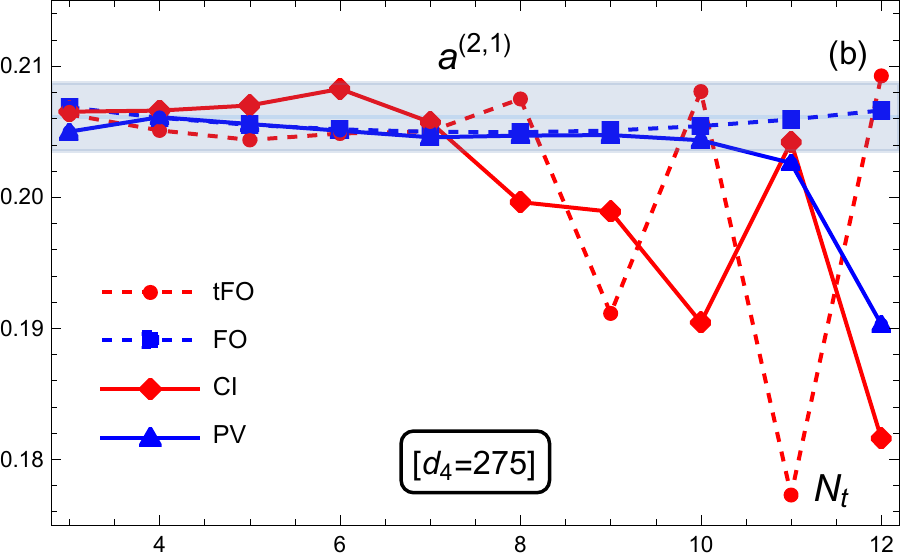}
\end{minipage}
\vspace{-0.2cm}
\caption{\footnotesize  (coloured online) The moment $a^{(2,0)}(\sm)$ (a) and $a^{(2,1)}(\sm)$ (b), as a function of the truncation index $N_t$, in the four considered approaches. At each $N_t$, the corresponding values of the parameters $\alpha_s$ and $\langle O_D \rangle$ obtained from the Borel-Laplace fit were used. The light blue band represents the experimental values (based on the ALEPH data).}
\label{a2javar}
\end{figure}

The extracted values for $\alpha_s$, with uncertainties from various sources given separately, are
\begingroup\color{black}
\bes
\label{BLresal}
\bea
\alpha_s(m_{\tau}^2)^{\rm (FO)} & = & 0.3228 \pm 0.0003({\rm exp})^{-0.0026}_{+0.0070}(\kappa)^{-0.0103}_{+0.0079}(d_4)^{+0.0081}_{-0.0057}(N_t)
\label{BLalFOa}
\\
& = &  0.3228^{+0.0134}_{-0.0121} \approx 0.323^{+0.013}_{-0.012},
\label{BLalFOb}
\\
\alpha_s(m_{\tau}^2)^{\rm ({\widetilde {\rm FO}})} & = & 0.3209 \pm 0.0003({\rm exp})^{-0.0038}_{+0.0201}(\kappa)^{-0.0039}_{+0.0047}(d_4)^{+0.0293}_{-0.0084}(N_t)
\label{BLaltFOa}
\\
& = &  0.3209^{+0.0359}_{-0.0100} \approx 0.321^{+0.036}_{-0.010},
\label{BLaltFOb}
\\
\alpha_s(m_{\tau}^2)^{\rm (PV)} & = & 0.3269 \pm 0.0003({\rm exp})^{+0.0007}_{+0.0102}(\kappa)^{-0.0064}_{+0.0155}(d_4)^{+0.0092}_{-0.0006}(N_t)^{+0.0167}_{-0.0067}({\rm amb})
\label{BLalPVa}
\\
& = &  0.3269^{+0.0266}_{-0.0093} \approx 0.327^{+0.027}_{-0.009}.
\label{BLalPVb}
\\
\alpha_s(m_{\tau}^2)^{\rm (CI)} & = & 0.3488 \pm 0.0005({\rm exp})^{+0.0078}_{+0.0004}(\kappa) \pm 0.0000(d_4)^{-0.0027}_{+0.0119}(N_t)
\label{BLalCIa}
\\
& = &  0.3488^{+0.0142}_{-0.0028} \approx 0.349^{+0.014}_{-0.003},
\label{BLalCIb}
\eea
\ees
\endgroup
The central values were extracted for the truncation index $N_t=8,5,8,4$ for the methods FO, ${\widetilde {\rm FO}}$, PV and CI, respectively, cf.~Table \ref{tabBL},\footnote{In Tables IV and V of Ref.~\cite{EPJ21}, the values of the condensates $\langle O_D \rangle$ were written in units $10^{-3} \ {\rm GeV}^D$.} and these values of $N_t$ were obtained by looking for the local stability of the resulting FESRs $a^{(2,j)}(\sm)$ ($j=0,1$) under variation of $N_t$ as explained earlier (see also Figs.~\ref{a2javar} later below). In the results (\ref{BLresal}), at the symbol '($\kappa$)' is the variation when the renormalisation scale parameter $\kappa$ ($ \equiv \mu^2/Q^2$) is vared from $\kappa=1$ up to $\kappa_{\rm max}=2$ and \textcolor{black}{down to $\kappa_{\rm min}=2/3$}, respectively.\footnote{\textcolor{black}{If we decreased $\kappa$ to $\kappa_{\rm min}=1/2$, the variation of the results would increase because of the vicinity of the Landau singularities of the ($\MSbar$) pQCD coupling $a(0.5 \sm e^{i \phi})$ in such a case ($0.5 \sm =1.4 \ {\rm GeV}^2$ is quite low; the Landau pole is at about $0.4 \ {\rm GeV}^2$), and the $\kappa$-dependence would become (artificially) the dominant uncertainty.}} At the symbol '($N_t$)' is the maximal variation when the truncation number is varied around its central value $N_t$: \textcolor{black}{in the cases of FO, ${\widetilde {\rm FO}}$, and PV the variation was in the interval $N_t=8^{+2}_{-3}$, $5 \pm 2$, $8^{+2}_{-3}$, respectively (thus the case $N_t=5$ that is independent of the renormalon model is always included);} in the case of CI, $N_t = 4^{+2}_{-1}$, i.e., $N_t=2$ was not considered, for being an extreme truncation). At the symbol '($d_4$)' is the variation when the coefficient $d_4$ varies according to Eq.~(\ref{d4est}) [cf.~also Table \ref{tabrenmod1}]. At the symbol '(exp)', the variations are (rough) estimates of the experimental uncertainties, and were obtained by the method explained in \cite{EPJ21}.\footnote{Cf.~discussion after Eqs.~(58) of Ref.~\cite{EPJ21}.} In the PV method, there is an additional (fourth) source of theoretical uncertainty '(amb)', which is  an estimate of uncertainty due to the Borel integration ambiguity for the Adler function \cite{EPJ21}.

In the results Eqs.~(\ref{BLresal}) and Table \ref{tabBL}, the total uncertainties were obtained by adding the mentioned various uncertainties in quadrature. We see from Eqs.~(\ref{BLresal}) that the main sources of uncertainties are theoretical, especially `$(d_4)$' and '$(\kappa)$'.

In Figs.~\ref{a2javar}(a)-(b) we present the (double-pinched) FESR momenta $a^{(2,0)}(\sm)$ and $a^{(2,1)}(\sm)$, Eqs.~(\ref{a2n})-(\ref{a2021}),  at each truncation index $N_t$, for the four evaluation methods (FO, ${\widetilde {\rm FO}}$, CI, PT).
For each $N_t$ and each method, we use the corresponding values of the parameters $\alpha_s$, $\langle O_4 \rangle$,  $\langle O_6^{(1)} \rangle$ and $\langle O_6^{(2)} \rangle$ obtained from the described (double-pinched) Borel-Laplace fits. From these Figures we can deduce that the (relatively) most stable values under the variation of $N_t$ are $N_t=8, 5, 4, 6$ for FO, ${\widetilde {\rm FO}}$, CI, PT, respectively. For this reason, we chose these values of $N_t$ as the central values for the respective methods in Eqs.~(\ref{BLresal}). We can see that in general, and for a reasonably wide range of $N_t$, the resulting predicted values of $a^{(2,0)}(\sm)$ and $a^{(2,1)}(\sm)$ are well compatible with the experimental results. 

\begin{figure}[htb] 
\begin{minipage}[b]{.49\linewidth}
  \centering\includegraphics[width=85mm]{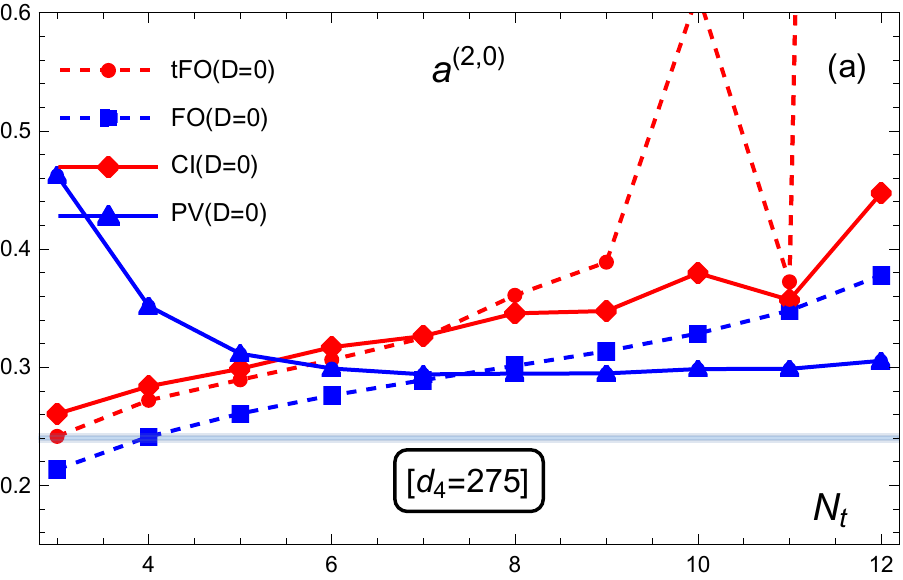}
  \end{minipage} 
\begin{minipage}[b]{.49\linewidth}
  \centering\includegraphics[width=85mm]{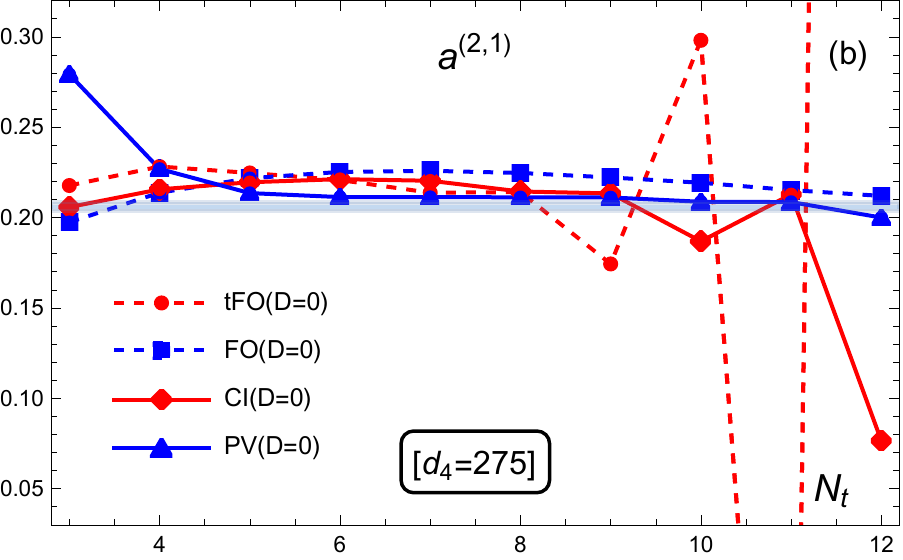}
\end{minipage}
\vspace{-0.2cm}
\caption{\footnotesize  (coloured online) The moment $a^{(2,0)}(\sm)_{D=0}$ (a) and $a^{(2,1)}(\sm)_{D=0}$ (b), as a function of the truncation index $N_t$, in the four considered approaches. For all $N_t$, the corresponding value of the parameter $\alpha_s$ was kept fixed for each method (cf.~the text for discussion).}
\label{a2jafixed}
\end{figure}

One may ask how the momenta $a^{(2,0)}(\sm)$ and $a^{(2,1)}(\sm)$ behave under variation of $N_t$ when the parameters $\alpha_s$, $\langle O_4 \rangle$,  $\langle O_6^{(1)} \rangle$ and $\langle O_6^{(2)} \rangle$ are not varied but kept fixed. In that case, only the $D=0$ contribution varies with $N_t$, while $D=4, 6$ contributions become $N_t$-independent. In Figs.~\ref{a2jafixed}(a)-(b) we present these results, using for $\alpha_s$ the central value of each method, i.e., the corresponding central values in Table \ref{tabBL}.
We can see that, in contrast to the results of Figs.~\ref{a2javar}(a)-(b), now the predictions are much more unstable under the variation of the truncation index $N_t$, indicating the importance of the inclusion of the OPE contributions $D=4$ and $D=6$ in the Borel-Laplace fit analysis (at each $N_t$).

\begingroup\color{black}
At the end of this Section, we wish to address the following question. In the considered $\MSbar$ scheme we know the $\beta$-function (\ref{RGE}) only up to the five-loop level (i.e., up to the coefficient $\beta_4$ \cite{5lMSbarbeta}), but we nonetheless use in the FO and PV approaches the relatively high truncation index $N_t=8$. In principle, due to the connections (\ref{dntdk}), the Adler function perturbation coefficients $d_n$ generated from the perturbation coefficients $\td_{n-s}$ of ${\cal B}[\td](u)$ involve the coefficients $k_s(n+1-s)$ ($s=1,\ldots,n-1$) which in turn are functions of $\beta_j$ ($j=1,\ldots,n-1$) \cite{renmod}. This implies that $d_6$ involves $\beta_5$ and $d_7$ (at $a^{N_t} = a^8$) involves $\beta_5$ and $\beta_6$ coefficients which are taken to be zero here. Such an effect of truncation of the $\beta$-function, i.e, setting $\beta_5=\beta_6=0$, in principle changes the coefficients $d_n$ ($n \geq 6$) deduced from the Borel transform ${\cal B}[\td](u)$ Eq.~(\ref{Btd5P}) of the renormalon-motivated model. In \cite{renmod}, these effects were investigated in the case of the so-called Lambert MiniMOM (LMM) scheme, and it was shown there that the truncation of the LMM scheme at four-loop level ($\beta_j=0$ for $j \geq 4$) gives numerically very similar results for the relevant parameters ${\cal C}_{j,k}^{(D)}$ and ratios of residues $d^{X}_{p,k}/{\td^{X}_{p,k}}$, leading to similar expressions for the Borel transform ${\cal B}[d](u)$ of the Adler function in the two cases and thus to similar values of $d_n$ ($n \geq 5$).\footnote{\textcolor{black}{For details, the notations and the results, we refer to \cite{renmod}, and in particular to Tables I and II there (LMM and TLMM cases). For the definition of the LMM $\beta$-function (of Pad\'e-type), we refer to Eq.~(8) of \cite{3dAQCD}.}} For these reasons, we believe that in the renormalon-motivated model considered here (in the $\MSbar$ scheme), the values of the coefficients $d_n$ in FO and $\delta d_n$ in PV approach ($n=6,7$) are not affected significantly by the setting $\beta_5=\beta_6=0$.  
\endgroup

\section{Final results and conclusions}
\label{sec:concl}

We applied double-pinched Borel-Laplace sum rules to the V+A channel semihadronic strangeless $\tau$ decay data of ALEPH. We used for $D=0$ contribution to the Adler function the renormalon-motivated extension \cite{renmod} for the higher order terms. We applied four different methods of evaluation (FO, ${\widetilde {\rm FO}}$, CI, PV). The optimal truncation index $N_t$ (of the $D=0$ Adler contribution) in the Borel-Laplace sum rules was fixed in such a way that the predicted FESR momenta $a^{(2,0)}$ and $a^{(2,1)}$ showed local stability under the variation of $N_t$.

As argued in detail in our previous work, the FO, ${\widetilde {\rm FO}}$ and PV methods of evaluation of the sum rules lead to truncated series\footnote{In the PV case, the truncated series refers to the series $\delta d(\sm e^{i \phi})^{[N_t]}_{D=0}$ in Eq.~(\ref{PV2}) which is free from the leading renormalon singularities and gives the corresponding contour integrals (sum rules) also free from the leading renormalon singularities.} that have the leading renormalon contribution of the Adler function (double UV renormalon pole at $u=+1$) suppressed in them; and that the CI method in the sum rules does not have this property.\footnote{These arguments, presented in a very general way in \cite{EPJ21} (especially Appendix A there), are also compatible with somewhat related arguments presented in \cite{BJ,BJ2,BoiOl}. A construction of Borel transforms of CI sum rules was proposed and investigated in \cite{HoangR} (cf.~also \cite{HoangR2}).}  We refer for details on these points to \cite{EPJ21}, and especially Appendix A there.

The four different methods (FO, ${\widetilde {\rm FO}}$, CI, PV) give the main results presented, for each method, in Eqs.~(\ref{BLresal}) and Table \ref{tabBL}. On the grounds argued above, we exclude the CI methods from our average, and the arithmetic average of the three methods (FO, ${\widetilde {\rm FO}}$ and PV) then gives
\begingroup\color{black}
\bes
\label{3av}
\bea
\alpha_s(m_{\tau}^2) &=& 0.3235^{+0.0138}_{-0.0126} \qquad ({\rm FO}+{\widetilde {\rm FO}}+{\rm PV}) 
\label{3ava} \\
\Rightarrow \;
\alpha_s(M_Z^2) &=& 0.1191 \pm 0.0016.  
\label{3avb} \eea \ees
We regard this as the central result of our analysis. The uncertainties $^{+0.0138}_{-0.0126}$ in Eq.~(\ref{3ava}) were obtained by adding in quadrature the largest deviation between the average value $0.3235$ and the central values of the three methods Eqs.~(\ref{BLresal}) ($\pm 0.0034$), and the uncertainties of the method which gives the smallest uncertainties among the three methods ($^{+0.0134}_{-0.0121}$) [cf.~Eq.~(\ref{BLalFOb})], similar to the reasoning in Refs.~\cite{Pich,EPJ21}. 
\endgroup

In Table \ref{tabrescomp} we present, for comparison, the values of $\alpha_s(m_{\tau}^2)$ which were extracted from the ALEPH $\tau$-decay data by various groups, who used various sum rules and various methods of evaluation.
\begin{table}
  \caption{The values of $\alpha_s(m_{\tau}^2)$, extracted by various groups applying sum rules and various methods to the ALEPH $\tau$-decay data. \textcolor{black}{BL stands for Borel-Laplace sum rules, and DV for a Duality Violation model in sum rules.}}
 \label{tabrescomp}
\begin{ruledtabular}
\begin{tabular}{l|l|lll|l}
group &  sum rule & FO & CI & PV & average \\
\hline
Baikov et al.~\cite{BCK} & $a^{(2,1)}=r_{\tau}$ & $0.322 \pm 0.020$ & $0.342 \pm 0.011$ & --- & $0.332 \pm 0.016$ \\
Beneke \& Jamin \cite{BJ} & $a^{(2,1)}=r_{\tau}$ & $0.320^{+0.012}_{-0.007}$ & --- & $0.316 \pm 0.006$ & $0.318 \pm 0.006$ \\
Caprini \cite{Caprini2020} & $a^{(2,1)}=r_{\tau}$ &  --- & --- & $0.314 \pm 0.006$ &  $0.314 \pm 0.006$ \\
Davier et al.~\cite{ALEPHfin} & $a^{(i,j)}$ & $0.324$ & $0.341 \pm 0.008$ & --- & $0.332 \pm 0.012$ \\
Pich \& R.-S.~\cite{Pich}   &  $a^{(i,j)}$     & $0.320 \pm 0.012$ &  $0.335 \pm 0.013$ & --- & $0.328 \pm 0.013$  \\
Boito et al.~\cite{Bo2015} & DV in $a^{(i,j)}$ & $0.296 \pm 0.010$ & $0.310 \pm 0.014$ & --- & $0.303 \pm 0.012$ \\
our previous work \cite{EPJ21}  & BL & $0.308 \pm 0.007$ & $0.335^{+0.010}_{-0.007}$ & $0.316^{+0.008}_{-0.006}$ & $0.312 \pm 0.007$ (FO+PV) \\
this work  & BL &
\begingroup\color{black}
$0.323^{+0.013}_{-0.012}$
\endgroup
&
$0.349^{+0.014}_{-0.003}$
&
\begingroup\color{black}
$0.327^{+0.027}_{-0.009}$
\endgroup
&
\begingroup\color{black}
$0.324 \pm 0.013$
(FO+${\widetilde {\rm FO}}$+PV) 
\endgroup
\end{tabular}
\end{ruledtabular}
\end{table}
We refer to \cite{EPJ21} for discussion and comparison of various of the approaches and results presented in the Table. None of those works used OPE in the form (\ref{DOPE}) but rather in the form (\ref{DOPE1}).

If, on the other hand, we included the CI method results in the average, the value of the coupling would significantly increase
\begingroup\color{black}
\bes
\label{4av}
\bea
\alpha_s(m_{\tau}^2) &=& 0.3299^{+0.0232}_{-0.0225} \qquad ({\rm FO}+{\widetilde {\rm FO}}+{\rm PV}+{\rm CI}) 
\label{4ava} \\
\Rightarrow \;
\alpha_s(M_Z^2) &=& 0.1199^{+0.0026}_{-0.0028}.  
\label{4avb} \eea \ees
\endgroup

\textcolor{black}{If we used, instead of the estimate (\ref{d4est}) for $d_4$ the higher estimate $d_4=338.19 \pm 63.19$, then the repetition of the described analysis gives, instead of the result (\ref{3av}), somewhat lower values $\alpha_s(m^2_{\tau})=0.3198^{+0.0139}_{-0.0148}$ and $\alpha_s(M_Z^2) = 0.1187^{+0.0016}_{-0.0019}$. 
We notice that the latter result is significantly higher than the corresponding result when the more naive OPE (\ref{DOPE1}) (truncated at  $D \equiv 2 k = 8$) is used for the Adler function \cite{EPJ21} (and $d_4=338.19 \pm 63.19$): $\alpha_s(m_{\tau}^2)=0.3116 \pm 0.0073$ [$\alpha_s(M_Z^2)=0.1176 \pm 0.0010$]. Therefore, we conclude that the values of the extracted coupling are numerically significantly increased when we use the improved form (\ref{DOPE}) of OPE (correspondingly truncated at $D \equiv 2 k = 6$), instead of the traditionally used (truncated) OPE form (\ref{DOPE1}): $\delta \alpha_s(m^2_{\tau}) \approx +0.008$. Further, the values are further increased (but somewhat less) when we decrease the estimate $d_4=338.19 \pm 63.19$ to  $d_4=275. \pm 63.19$ [Eq.~(\ref{d4est})]: $\delta \alpha_s(m^2_{\tau}) \approx +0.004$.}

\textcolor{black}{Since our results have relatively high truncation index $N_t=8$ (where $a^{N_t}$ is the power where truncation is made) for the FOPT and PV approaches, one may wonder how much the obtained results depend on the specific renormalon-motivated Adler function. The FOPT is independent of the renormalon-motivated extension for $N_t=5$ [the coefficient $d_4$ is taken according to Eq.~(\ref{d4est})], and this is true to a large degree also for the PV approach. As can be seen from the results (\ref{BLalFOa}) and  (\ref{BLalPVa}), the extracted values of the coupling change only little if $N_t=5$ is taken instead of $N_t=8$:  the central value of $\alpha_s(m^2_{\tau})$ changes from $0.3228$ to $0.3171$ in the FOPT case, and from $0.3269$ to $0.3277$ in the PV case; the average central value $\alpha_s(m^2_{\tau})=0.3235$ Eq.~(\ref{3av}) changes to $\alpha_s(m^2_{\tau})=0.3219$ when $N_t=5$ in all three methods is taken [and $\alpha_s(M_Z^2)$ changes from $0.1191$ to $0.1189$]. Therefore, we can conclude that the influence of the extension of the Adler function beyond the order $a^5$ (by our renormalon-motivated model) does not have a numerically important role in the determination of $\alpha_s$. The case $N_t=5$ is properly included in the uncertainties of the results Eqs.~(\ref{BLresal}).}

\textcolor{black}{The effects of the quark hadron duality violations (DVs) may be important for specific moments in the sum rules, as argued in \cite{Bo2017,Bo2021} where a DV model \cite{Cata} was used. On the other hand, we extracted our results from the Borel-Laplace sum rules whose weight function Eq.~(\ref{gM2}) is suppressed at the Minkowskian point $Q^2=-\sm$ through the double-pinching factor $(1 + Q^2/\sm)^2$ and, in addition, the exponential factor $\exp(Q^2/M^2)$. The exponential factor suppression is especially strong and extended around the mentioned Minkowskian point when $|M^2|$ is small ($|M^2| \approx 1 \ {\rm GeV}^2$), i.e., at those values of $|M^2|$ where the nonperturbative effects ($D =4, 6$ terms) in the Borel-Laplace sum rule are large. This indicates that the applied Borel-Laplace sum rules are designed to suppress the DV effects more strongly when these DV effects are stronger.}

The results were obtained by an analysis based on programs written by us in Mathematica that are freely available from \cite{prgs}.

\begin{acknowledgments}
This work was supported in part by FONDECYT (Chile) Grants No.~1200189 and No.~1180344 and ANID Fellowship No.~21211716.
\end{acknowledgments}

\appendix

\section{General explicit expressions for Borel-Laplace and FESR momenta $a^{(2,n)}$}
\label{app:genexpr}

The general explicit expression for the contribution of the dimension $D=2 k$ ($k =2,3,4,\ldots$) of the theoretical Borel-Laplace sum rule [Eq.~(\ref{Bth})] is given in Eqs.~(\ref{BthD}). In these expressions, for the parts involving only $\langle O_{2 k} \rangle^{(1)} + \langle O_{2 k} \rangle^{(2)}/a(\sm)$, the following integrals were used:
\be
\frac{1}{2 \pi} \int_{-\pi}^{+\pi} d \phi \; \exp(- i s \phi) \exp \left( A e^{i \phi} \right) = \frac{A^s}{s!},
\label{intexp} \ee
where $s=0,1,2,\ldots$ and $A$ is a complex number ($A=\sm/M^2$). These expressions were explained in \cite{EPJ21} (App.~A.2 there). On the other hand, for $s$ negative integer ($s=-1,-2,\ldots$) these integrals are zero.

\begin{figure}[htb] 
  \centering\includegraphics[width=60mm]{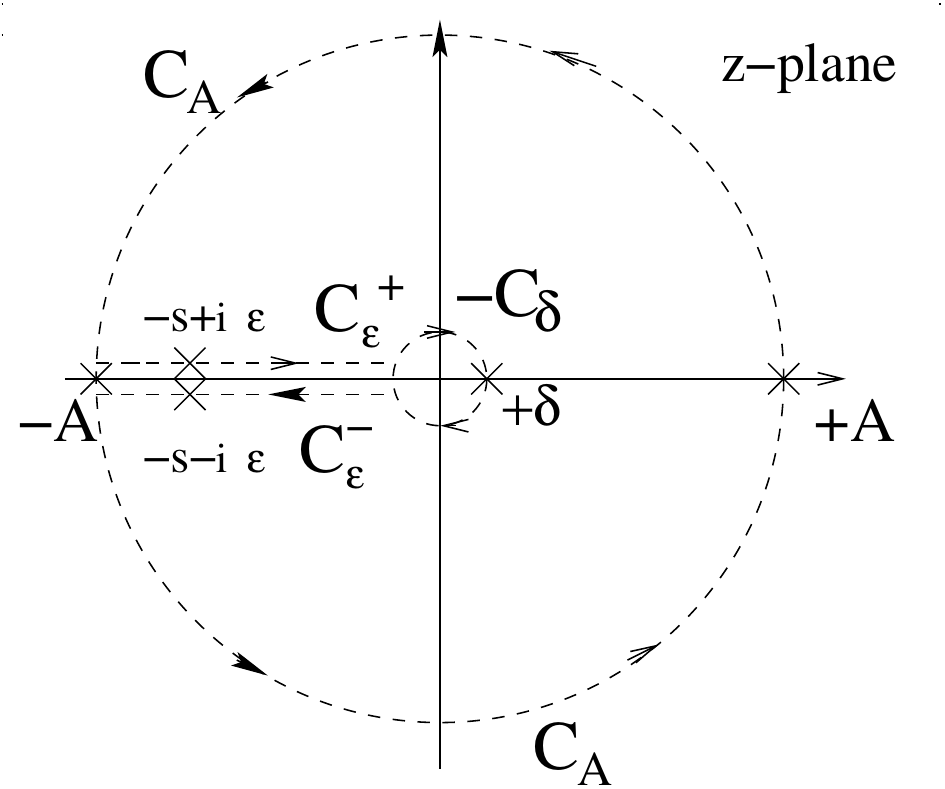}
\vspace{-0.2cm}
\caption{\footnotesize  The closed integration path in the complex $z$-plane for the integral Eq.~(\ref{Isa}): $C_A$ is the (counterclockwise) circular path of radius $A$ ($A>0$); $C_{\varepsilon}^{\pm}$ are the linear paths parallel to the negative axis ($z=-s \pm i \varepsilon$, $-A \leq s \leq -\delta$; $-C_{\delta}$ is the (clockwise) circular path of radius $\delta$. The limit $\varepsilon \ll \delta \to +0$ is taken, i.e., first $\varepsilon \to +0$ and then $\delta \to +0$.}
\label{contIs}
\end{figure}

The parts proportional to $\langle O_{2 k} \rangle^{(2)} \beta_0$ involve the integrals $J_s(A)$ given in Eq.~(\ref{Js}). The change of variable $z = A e^{i \phi}$ ($s$ is a nonnegative integer) gives
\be
J_s(A) = \frac{(-i)}{2 \pi} A^s \oint_{C_{|A|}} \frac{d z}{z^{s+1}} e^z \ln \left( \frac{z}{A} \right),
\label{Jsb} \ee
where the integration in the complex $z$-plane runs along the circle of radius $|A|$, $z=A e^{i \phi}$.
If we assume first that $A > 0$, then the application of the Cauchy theorem to the integral 
\be
\oint_{C} \frac{d z}{z^{s+1}} e^z \ln z = 0
\label{Isa} \ee
along the closed contour $C = C_{A} + C_{\varepsilon}^{+} + C_{\varepsilon}^{-} - C_{\delta}$ depicted in Fig.~\ref{contIs},
and taking the limit $0 < \varepsilon \ll \delta \to +0$, gives
\bea
\oint_{C_{A}} \frac{d z}{z^{s+1}} e^z \ln z & = &
- \int_{C_{\varepsilon}^{+} + C_{\varepsilon}^{-} - C_{\delta}} \frac{d z}{z^{s+1}} e^z \ln z
\nonumber\\
& = & \frac{2 \pi i}{s!} \ln A + \frac{2 \pi i}{A^s} (-1)^s \sum_{\substack{k=0\\(k \not=s)}}^{\infty} \frac{ (-1)^k A^k}{(k-s) k!}.
\label{Isb} \eea
When this is combined with the integral
\be
\oint_{C_{A}} \frac{d z}{z^{s+1}} e^z (-\ln A) = - \frac{2 \pi i}{s!} \ln A,
\label{Isc} \ee
we obtain
\bea
J_s(A) &=& A^s \sum_{\substack{\ell=-s\\ \ell \not=0}}^{\infty} \frac{ (-1)^{\ell} A^{\ell}}{\ell (\ell + s)!}. 
\label{Jsres} \eea
 When $s=1,2,3$, the following explicit expressions are obtained from (\ref{Jsres})
\bes \label{J123}
\bea
J_1(A) & = & e^{-A} - A \left[ -1 + \gamma_{\rm E} + \Gamma(0,A) + \ln A \right]
\label{J1expl} \\
J_2(A) & = & \frac{1}{2}(-1 +A) e^{-A} - \frac{A^2}{4} \left[ -3  + 2 \gamma_{\rm E} + 2 \Gamma(0,A) + 2 \ln A \right]
\label{J2expl} \\
J_3(A) & = & \frac{1}{6}(2 -A +A^2) e^{-A} - \frac{A^3}{36} \left[ -11  + 6 \gamma_{\rm E} + 6 \Gamma(0,A) + 6 \ln A \right],
\label{J3expl} \eea \ees
where $\Gamma(a,A)$ is the incomplete Gamma function (we have here $a=0$), and $\gamma_{\rm E}$ is the Euler-Mascheroni constant ($\gamma_{\rm E} \approx 0.5772$).
When $A$ is not positive real but complex, these same formulas remain valid by complex continuation in $A$. It can be checked by numerical evaluation (integration) of the integrals $J_s(A)$ Eq.~(\ref{Js}) for complex values of $A$ that the explicit expressions (\ref{J123}) are valid.

The general expression for the contribution of the dimension $D= 2 k$ ($k \geq 2$) operators to the (double-pinched) FESR momenta $a^{(2,n)}(\sm)$ ($n \geq 0$) is
\bea
a^{(2,n)}(\sm)_{D= 2 k} &=& \frac{1}{2 \pi} \int_{-\pi}^{+\pi} d \phi \; G^{(2,n)} \left( \sm e^{i \phi} \right) \frac{2 \pi^2 k}{\sm^k} e^{ - i k \phi} \left[   \langle O^{(1)}_{2 k} \rangle +  \langle O^{(2)}_{2 k} \rangle \left( \frac{1}{a(\sm)} + i \beta_0 \phi \right) \right]
    \label{a2nint}
\eea
where we took one-loop running of $1/a(Q^2)$ on the contour $Q^2 = \sm e^{i \phi}$ around $\sm$, Eq.~(\ref{O2arun}). Evaluation of this integral then gives
 \bea
 a^{(2,n)}(\sm)_{D= 2 k} & = &  \frac{2 \pi^2 k}{\sm^k} {\Bigg \{} \left[ \langle O^{(1)}_{2 k} \rangle + \frac{\langle O^{(2)}_{2 k} \rangle}{a(\sm)} \right] \left[ \left( \frac{n+3}{n+1} \right) \delta_{k,n+2} +  \delta_{k,n+3} \right] (-1)^n
\nonumber\\
&& +  \langle O^{(2)}_{2 k} \rangle \beta_0 (-1)^k \left[ \left( \frac{n+3}{n+1} \right) \left( \frac{1}{k-1} - \frac{(1- \delta_{k,n+2})}{(k-n-2)} \right) -
  \left( \frac{1}{k} - \frac{(1- \delta_{k,n+3})}{(k-n-3)} \right) \right]
    {\Bigg \}}.
    \label{a2nDexpl} \eea

\end{document}